\documentclass[a4paper,11pt,twoside]{article}

\usepackage{geometry} 
\usepackage{amsmath}
\usepackage{amssymb}
\usepackage{color}
\usepackage{graphicx}
\usepackage{cite}

\newcommand{\GeV}{\,\mathrm{GeV}}

\newcommand{\fracwithdelims}[4]{\left#1 \frac{#3}{#4} \right#2}
\newcommand{\ord}[1]{\mathcal{O}\left( #1 \right)}

\newcommand{\vev}[1]{\left\langle #1\right\rangle}
\newcommand{\VeV}[2]{#1\langle #2 #1\rangle} 
\newcommand{\dm}[1]{{\Delta m^2_{#1}}}

\newcommand{\Eq}[1]{Eq.~(\ref{eq:#1})}
\newcommand{\eq}[1]{eq.~(\ref{eq:#1})}

\newcommand{\eqs}[1]{eqs.~(\ref{eq:#1})}

\newlength{\myem}
\newcommand{\sep}[1]{#1}
\newcounter{mysubequation}[equation]
\renewcommand{\themysubequation}{\alph{mysubequation}}
\newcommand{\mytag}{\stepcounter{mysubequation}%
\tag{\theequation\protect\sep{\themysubequation}}}
\newcommand{\globallabel}[1]{\refstepcounter{equation}\label{#1}}

\newcommand{\GPS}{G_{\text{PS}}}

\newcommand*{\be}{\begin{equation}}
\newcommand*{\ee}{\end{equation}}
\newcommand*{\bea}{\begin{eqnarray}}
\newcommand*{\eea}{\end{eqnarray}}

\newcommand{\SISSA}{SISSA/ISAS and INFN, I--34013 Trieste, Italy}

\newcommand{\preprintnumber}{%
SISSA--73/2008/EP}
 
\newcommand{\titletext}{Consequences of a unified, anarchical model \\ of fermion masses and mixings} 
\newcommand{\authortext}{L. Calibbi, L. Ferretti, A. Romanino, R. Ziegler
\medskip\\\em\normalsize 
\SISSA}
\newcommand{\abstracttext}{%
We show that most features of the mass and mixing pattern of the second and third SM fermion families can be accounted for without making use of flavour symmetries or other types of flavour dynamics. We discuss the implications for flavour phenomenology, in particular for the $\tau\to\mu\gamma$ decay rate, and comment on LFV effects at colliders. We show that the model can be embedded in a full SO(10) supersymmetric GUT in 5 dimensions that preserves the successful MSSM gauge coupling unification prediction for $\alpha_s$. Interesting features of this embedding are i) the connection of one of the hierarchy parameters with the strong coupling assumption, ii) the absence of KK threshold effects on the $\alpha_s$ prediction at one loop, and iii) the shift of the GUT scale up to about $10^{17}\GeV$. Proton decay is under control, also due to the larger GUT scale. A large atmospheric angle for normal hierarchical neutrinos is obtained in an unusual way.}

\title{
\normalsize
\begin{tabular}[t]{l}
\end{tabular}
\hspace*{\fill}
\begin{tabular}[t]{l}\preprintnumber\end{tabular}
\vspace{3\baselineskip}\\\Large\bfseries\titletext\bigskip}
\author{\begin{minipage}[t]{0.8\textwidth}
\normalsize\centering\authortext
\end{minipage}}
\date{}

\begin{document}

\bigskip
\maketitle
\begin{abstract}\normalsize\noindent
\abstracttext
\end{abstract}\normalsize\vspace{\baselineskip}


\section{Introduction}

Different approaches have been used so far to explain the peculiar pattern of fermion masses and mixings and in particular to account for the widely different patterns observed in the charged and neutral fermion sectors. 

While the experimental determination of neutrino masses and mixings is relatively recent, the first charged fermion mass was measured more than a century ago. As a consequence, most approaches, most notably the one based on flavour symmetries, have been developed in connection to the charged fermion sector and have been later extended to neutrinos and lepton mixing. This is also because charged fermion masses may be related to the origin of flavour in a more direct way than neutrinos, as the latter are likely to get their masses from lepton number violation at large scales through some kind of see-saw mechanism. 

Following~\cite{Ferretti:2006df}, in this paper we will consider an ``anarchical'' approach that does not make use of any flavour symmetry \cite{Dobrescu&Barr} or other dynamical mechanisms and can be considered to be inspired by the neutrino sector. It is well known that the structure of masses and mixings of the second and third neutrino families can be accounted for in terms of the dominant exchange of a single right-handed neutrino with similar, possibly large couplings to both the second and third light neutrinos\cite{King}. This leads to a normal-hierarchical pattern of light neutrinos. As the couplings can be taken to be random order one numbers, this mechanism does not require any special dynamics controlling the sizes of the couplings of the second and third families. Still, the neutrino spectrum ends up being hierarchical, for the simple reason that the exchange of a single singlet neutrino can only give mass to one light neutrino. 

Is it possible that a similar mechanism is at work in the charged fermion sector, with the structure of masses and mixings of the second and third charged fermion families accounted for by the dominant exchange of a single set of messengers, with unconstrained, anarchical, $\ord{1}$ couplings? On the one hand, the single messenger mechanism can in principle be exported, since many theories of flavour, including those based on flavour symmetries, assume the existence of a heavy sector of messengers. On the other hand, the extension to the quark sector might look non-trivial, since the random $\ord{1}$ couplings lead in the neutrino sector to large mixing angles that we do not observe in the quark sector. This potential problem, however, does not arise if the messengers are left-handed, as observed in \cite{Ferretti:2006df}. In this case, in fact, the large rotations induced in the up and down quark sector are approximately equal and compensate each other, leading to small mixing in the CKM matrix. Starting from this observation, a Pati-Salam (PS) model was developed that accounts for most features of the masses and mixings of the second and third families and for the main qualitative features associated with the first family. While some specific flavour dynamics may need to be invoked for a quantitative account of the first family, it is impressive that so many other features do not actually need a flavour symmetry or other dynamics to be explained. 

In this paper, we show that the approach illustrated in \cite{Ferretti:2006df} can be embedded in a unified theory, more precisely in a SO(10) supersymmetric model in five dimensions. We also discuss the implications for flavour changing neutral current (FCNC) and lepton flavour violating (LFV) processes. We begin in Section~\ref{sec:lowE} illustrating the pattern of fermion masses and mixings emerging from the unified theory in the context of a simple, effective PS theory in four dimensions. In Section~\ref{sec:FCNC} we illustrate the peculiar FCNC and LFV effects associated with new physics near the GUT scale characterized by large couplings with both the second and third family. In Section~\ref{sec:SO(10)} we show an example of embedding in SO(10). In most of the paper we concentrate on the second and third family masses and mixings, whose features do not need flavour symmetries. In Section~\ref{sec:first}, before concluding, we sketch a possibility to fix the quantitative aspects related to the first family. For sake of clarity, we will give a short summary of the basic assumptions and achievements at the end of each section.

\section{Yukawa textures in the effective Pati-Salam theory}
\label{sec:lowE}

We begin by illustrating the model at scales lower than the compactification scale at which the unified structure emerges. As we will see in Section~\ref{sec:SO(10)}, SO(10) is broken at those scales to the Pati-Salam (PS) gauge group $\GPS = \text{SU(2)}_L\times\text{SU(2)}_R\times\text{SU(4)}_c$ and inherits two discrete symmetries, a parity $\mathbf{Z}_2$ and an $R$-parity $R_P$, from the full unified model. The chiral superfield content and the corresponding quantum numbers under the gauge and discrete symmetries are summarized in Table~\ref{tab:fields}. The first block contains the $\mathbf{Z}_2$-odd fields: the 3 light (in the unbroken $\mathbf{Z}_2$ limit) families $(f_i,f^c_i)$, $i=1,2,3$, the light Higgs $h$, and the $\mathbf{Z}_2$-breaking field $\phi$. The latter is assumed to be in the adjoint representation of SU(4)$_c$ as this provides the Georgi-Jarlskog factor 3 needed to account for the $\mu$--$s$ mass relation. The second block contains the messengers, in a single vectorlike family $(F,F_c)+(\bar F,\bar F_c)$. The third block contains the fields $F'_c+\bar F'_c$ and $X_c$ providing the necessary breaking of the Pati-Salam group. Finally, the field $\Sigma$ in the last column is needed to communicate the SU(2)$_R$ breaking provided by $F'_c+\bar F'_c$ to the messengers $F_c+\bar F_c$. The up and down components of those messengers need in fact to be different in order to account for $m_c/m_t\ll m_s/m_b$. The field content is thus rather economical. We will later add another few PS fields in order to preserve gauge coupling unification above $M_R$ and to take care of singlet neutrino masses. As we will see, the MSSM 1-loop gauge unification turns out to be exactly preserved. 

\begin{table}[htdp]
\begin{equation*}
\begin{array}{|c|cccc|cccc|ccc|c|}
\hline
& f_i & f^c_i & h & \phi & F & \bar F & F^c & \bar F^c & F'_c &  \bar F'_c & X_c & \Sigma \\ \hline
\text{SU(2)}_L & 2 & 1 & 2 & 1 & 2 & 2 & 1 & 1 & 1 & 1& 1 & 1 \\
\text{SU(2)}_R & 1 & 2 & 2 & 1 & 1 & 1 & 2 & 2 & 2 & 2 & 3 & 1 \\
\text{SU(4)}_c & 4 & \bar 4 &  1 & 15 & 4 & \bar 4 & \bar 4 & 4 & \bar 4 & 4 & 1 & 15 \\
\mathbf{Z}_2 & - & - & - & - & + & + & + & + & + & + & + & + \\
R_P & - & - & + & + & - & - & - & - & + & + & + & - \\
\hline
\end{array}
\end{equation*}
\caption{Chiral field content and quantum numbers under $G_{\text{PS}}$ and $\mathbf{Z}_2$}
\label{tab:fields}
\end{table}

The decomposition of the fields in Table~\ref{tab:fields} under the SM gauge group is the following (with hopefully self-explanatory notations): $F = (L,Q)$, $\bar F = (\bar L,\bar Q)$, $F^c = (L^c,Q^c)$, $\bar F^c = (\bar L^c, \bar Q^c)$, $L^c = (N^c,E^c)$, $Q^c = (U^c,D^c)$, $\bar L^c = (\bar N^c, \bar E^c)$, $\bar Q^c = (\bar U^c, \bar D^c)$, $\phi =  (A_\phi, T_ \phi , \bar T_ \phi , G_ \phi)$, where $A_\phi$ is a singlet, $T_\phi \sim(3,1,2/3)$, $\bar T_\phi \sim (\bar 3, 1, -2/3)$, $G_\phi \sim (8,1,0)$ under $\text{SU(3)}_c\times \text{SU(2)}_L \times\text{U(1)}_Y$ (all fields are properly normalized). Analogously for the other fields with the same quantum numbers under PS. We also denote $a\vev{X_c} = M_R\,\sigma_3$, $\vev{N'_c} = V_c$, $\vev{\bar N'_c} = \bar V_c$ ($|V_c|$ = $|\bar V_c|$ in the supersymmetric limit), $\vev{\phi} = M_L\, T_{B-L}$. 

At this effective level, the model is characterized by two scales, $M_R$ and $M_L$, with $M_R > M_L$, corresponding to the masses of the right-handed and left-handed messengers. The PS group is spontaneously broken to the SM group at the scale $M_R$. Up to some explicit mass terms, whose absence will be accounted for by the full theory in Section~\ref{sec:SO(10)}, the most general renormalizable superpotential for the fields in Table~\ref{tab:fields} is
\begin{equation}
\label{eq:RenLag}
W =
\lambda_i f^c_i F h + \lambda^c_i f_i F^c h +\alpha_i\phi f_i\bar F+\alpha^c_i\phi f^c_i\bar F^c + a\bar F^c X_c F^c + \bar\sigma_c \bar F'_c \Sigma F^c +\sigma_c\bar F^c\Sigma F'_c  + \frac{M_\Sigma}{2} \Sigma^2\ldots .
\end{equation}
All the couplings are assumed to be $\ord{1}$ and uncorrelated. The terms involving $R_P$-even fields only, providing the vevs of the fields $F'_c$, $\bar F'_c$, $X_c$, $\phi$ along the SM invariant directions, have been omitted, as they do not affect fermion masses directly. They will be discussed in Section~\ref{sec:SO(10)}. The $\mathbf{Z}_2$ conserving vevs lie near a single scale, $M_R$, the scale of the right-handed messengers $F^c$, $\bar F^c$. The $\mathbf{Z}_2$-breaking vev of $\phi$ lies at the smaller scale $M_L$, the scale of the left-handed messengers $F$, $\bar F$. The hierarchy of SM fermion masses originates from $\epsilon \equiv M_L/ M_R \ll 1$. This means that the hierarchy among different family masses originates from a hierarchy within a single family of messengers (which in turn is related to the hierarchy between two PS-breaking vevs). 

Non renormalizable terms in the superpotential involving vevs at the scale $M_R$ could give rise (or not) to mass terms comparable with the scale $M_L$. Such mass terms could be relevant for the left-handed messengers $F$, $\bar F$, which do not get a mass at the scale $M_R$. In the context of the full model, such a mass term does not arise\footnote{This represents a difference with respect to \cite{Ferretti:2006df}, where such a mass term was assumed to be present and to be along the $B-L$ direction.}. We assume that the Higgs field $h$ remains massless. This will be also accounted for in the full model, in terms of an $R$-symmetry. 

In order to identify the light (massless in the unbroken EW symmetry limit) MSSM fields, we plug the vevs in the superpotential in \eq{RenLag}. Since $R_P$ is not broken at this level, the $R_P$-even and $R_P$-odd heavy fields do not mix, and we can restrict our analysis to the $R_P$-odd fields. We choose a basis in flavour space such that $\lambda_{1,2} = \alpha_{1,2}=0$, $\lambda^c_1 = \alpha^c_1 = 0$. $\lambda_3$, $\alpha_3$, $\lambda^c_{2,3}$, $\alpha^c_{2,3}$, $M_R$, $v$, $V_c = \bar V_c$ can all be taken positive. The heavy mass terms turn out to be
\globallabel{eq:masses}
\begin{gather}
-\bar E^c \left[ M_R E^c - M_L(\alpha^c_3 e^c_3 +\alpha^c_2 e^c_2) \right]  -\alpha_3 M_L \bar L  l_3 \mytag \\
-\bar D^c \left[ M_R D^c +\frac{M_L}{3}(\alpha^c_3 d^c_3 +\alpha^c_2 d^c_2) \right] 
+ \alpha_3 \frac{M_L}{3}\bar Q q_3 \mytag  \\
+ \bar U^c \left[M_R U^c - \frac{\sigma_c}{\sqrt{2}}  V_c\bar T_\Sigma -\frac{M_L}{3} (\alpha^c_3 u^c_3 + \alpha^c_2 u^c_2) \right]
+T_\Sigma \left[M_\Sigma \bar T_\Sigma -\frac{\bar\sigma_c}{\sqrt{2}} 
\bar V_c U^c
\right] + \frac{M_\Sigma}{2} G^2_\Sigma \mytag  \\
+\bar N^c \left[ M_R N_c + M_L (\alpha^c_3 n^c_3 +\alpha^c_2 n^c_2) \right ]
+ \sqrt{\frac{3}{8}} \sigma_c V_c \bar N^c A_\Sigma + \sqrt{\frac{3}{8}} \bar\sigma_c 
\bar V_c N^c A_\Sigma +\frac{M_\Sigma}{2} A^2_\Sigma .  \mytag
\end{gather}
Because of the absence of $\bar LL$ and $\bar QQ$ mass terms, $L$ and $Q$ remain massless, while $l_3$ and $q_3$ get a mass together with $\bar L$ and $\bar Q$ from the vev of $\phi$. The light lepton and quark doublets are therefore $l'_3 = L$, $l'_{1,2} = l_{1,2}$, $q'_3 = Q$, $q'_{1,2} = q_{1,2}$ and the heavy ones are $L' = l_3$, $Q' = q_3$, $\bar L' = \bar L$, $\bar Q' = \bar Q$. The light SU(2)$_L$ singlets can be easily determined along the lines of \cite{Ferretti:2006df}. The $\mathbf{Z}_2$-even fields $D^c$, $U^c$, $E^c$ all contain a small, $\ord{\epsilon}$ light component, except $U^c$, whose light component ${u^c_2}'$ can be further suppressed because of the mixing with the color triplets $T_\Sigma$, $\bar T_\Sigma$ in $\Sigma$. A double $\epsilon$ suppression is obtained if 
\begin{equation}
\label{eq:rho}
\frac{M_\Sigma M^2_R}{M_L V^2_c} = \ord{1} .
\end{equation}
This double suppression then explains why $m_c/m_t\ll m_s/m_b$ \cite{Ferretti:2006df}. The implementation in Section~\ref{sec:SO(10)} will indeed give $M_\Sigma \sim M_R \sim \sqrt{\epsilon} V_c$~\footnote{In \cite{Ferretti:2006df} it was instead assumed that $V_c \sim M_R$ and $M_\Sigma \sim M_L$.}. 

The light fermion Yukawa matrices $Y^D$, $Y^E$, $Y^U$ (in right-left convention) are easily determined expressing the superpotential in terms of the massless fields. At the leading order in $\epsilon$ we find
\begin{equation}
\begin{gathered}
\label{eq:YDEU}
Y^D =  \begin{pmatrix}
0 & 0 & 0 \\
0 & -\alpha^c_2\lambda^c_2 \epsilon/3 & 0 \\
0 & -\alpha^c_3\lambda^c_2 \epsilon/3 & \lambda_3
\end{pmatrix} , \quad
Y^E =  \begin{pmatrix}
0 & 0 & 0 \\
0 & \alpha^c_2\lambda^c_2 \epsilon & 0 \\
0 & \alpha^c_3\lambda^c_2 \epsilon & \lambda_3
\end{pmatrix} , \\[2mm]
Y^U = \begin{pmatrix}
0 & 0 & 0 \\
0 & -(2/3)\alpha^c_2\lambda^c_2 \rho_u\epsilon^2 & 0 \\
0 & -(2/3)\alpha^c_3\lambda^c_2 \rho_u\epsilon^2 & \lambda_3
\end{pmatrix} ,
\end{gathered}
\end{equation}
where $\rho_u = (\sigma_c\bar\sigma_c)^{-1}(M_\Sigma M^2_R)/(M_L V^2_c)$ is an order one coefficient. The numerical value of $\epsilon$ turns out to be $\epsilon \approx 0.06\,\lambda_3/(\alpha^c_2\lambda^c_2)$, which implies $\rho_u \approx 1.9\, (\alpha^c_2\lambda^c_2/\lambda_3)$, indeed of order one. 

The model predicts the first family to be massless in the limit in which non-renormalizable corrections to $W$ are neglected and to be further suppressed by powers of the cutoff once those corrections are taken into account. Unlike the second and third families, the quantitative aspects related to the first family masses may need to be controlled by some flavour dynamics. An example is given in Section~\ref{sec:first}. 

Neutrino masses will be discussed in the context of the full model in Section~\ref{sec:neutrinos}.
\bigskip

To summarize, we illustrated a supersymmetric Pati-Salam model aiming at explaining the fermion masses and mixings for the 2nd and 3rd family. This model arises as the low-energy limit of a fully unified model in 5 dimensions, which will be presented in Section~\ref{sec:SO(10)}. The low energy effective limit of the full model is useful because it is simple, it contains most of the features relevant for the phenomenology, and it does not depend on the detailed implementation of the full unified model. The full model will also justify most of the assumptions made in this section, which are: 
\begin{itemize}
 \item Besides the Pati-Salam gauge symmetry there are 2 discrete symmetries, $\mathbf{Z}_2$ and $R_P$.
 \item All the allowed dimensionless couplings are present  and are $\ord{1}$; the structure and the scales of the mass terms will be accounted for by the full model in terms of an $R$-symmetry and a discrete $\mathbf{Z}_{24}$).
\item Only the $R_P$-even fields get a vev (the corresponding dynamics will be specified in the full model).
 \item We have 4 mass scales $M_R$, $M_L$, $M_\Sigma$, $V_c$ satisfying the relations
$$M_R \gg M_L, ~~\frac{M_\Sigma M^2_R}{M_L V^2_c} = \ord{1},$$ which account for the mass hierarchy in the down-quark  and charged lepton sector and the up-quark sector respectively (these scales will be dynamically justified in the full model). 
\end{itemize}

\section{FCNC and LFV}
\label{sec:FCNC}

In this section we discuss the implications of the effective model discussed in the previous section for Flavour Changing Neutral Current (FCNC) and Lepton Flavour Violation (LFV) effects. Most of the conclusions are independent of the details of the full unified implementation of the model and extend to the model in Section~\ref{sec:SO(10)}. 

In order to identify the peculiar effects associated to the basic ingredients of the model, we assume the supersymmetric soft terms to be universal at a scale $M_c > M_R$. More precisely, we assume a common value for the gaugino masses, $M_{1/2}$, for the sfermion masses, $m^2_0$, and for the trilinear terms, $A_0$, at that scale. As for the Higgs soft masses, we consider the possibility that they are independent, with an average value $(m^0_h)^2 = (m^2_{H_D} + m^2_{H_U})/2 = m^2_0$ and a splitting $4 m^2_X = m^2_{H_D} - m^2_{H_U}$. This is because their splitting $m^2_X$ can play an important role when $\tan\beta$ is large\footnote{In this regime $\lambda_t \approx \lambda_b \sim O(1)$ and the RGEs for $m_{H_D}^2$ and $m_{H_U}^2$ are similar. If $m_{H_D}^2 = m_{H_U}^2$ to start with, it is difficult to have at the same time $m_{H_U}^2+\mu^2$ negative enough to break the electroweak gauge symmetry and $m_{H_D}^2+\mu^2$ positive enough so that the squared pseudoscalar Higgs mass is positive. The splitting $m_X$ between the two Higgs soft terms can be given by additional contributions to the soft SUSY breaking scalar masses arising from the D-terms associated with the broken diagonal generators. These contributions are generated whenever the soft SUSY-breaking masses of the fields whose vacuum expectation values reduce the rank of the group are different \cite{Hagelin:1989ta}.} \cite{Ciafaloni:1995ad} and because the connection of Higgs and sfermion masses may not be straightforward\cite{GMSB,GauginoMSB}. For definiteness, we will always take $A_0 = 0$. Motivated by the unified model of Section~\ref{sec:SO(10)}, we take $M_c$ as determined by the unification condition ($M_c \sim 10^{16}\GeV$ in the region of parameter space in which threshold effects are favourable) and we also consider the limit $m^2_0 = m^2_X = 0$  in which the only sizable high-energy parameter is the common gaugino mass and the low-energy scalar soft masses and trilinear terms are generated by the RGE running. 

\subsection{The pattern of the RGE effects}

The interactions of MSSM fields at energy much higher than the electroweak scale leave their imprint on the sfermion soft masses through radiative effects. Well known examples are the lepton flavour violating effects induced by see-saw~\cite{Borzumati:1986qx} and GUT~\cite{Barbieri:1994pv} interactions. In the model described in the previous Section, the MSSM fields interact with the left-handed and right-handed flavour-messengers living at the scales $M_L$ and $M_R$ respectively. The peculiar feature of our model is that the corresponding Yukawa interactions (with all the three light families) are described by $\ord{1}$ couplings, not constrained to be small by any symmetry. We therefore expect the flavour-violating effects induced in the sfermion mass matrices to be sizable. 

In order to get an analytical understanding of the effects, we write the off-diagonal sfermion mass terms induced by the extra-interactions at $M_L$ and $M_R$ in the leading logarithm approximation. Also, we use the same basis in flavour space that leads to the textures in \eq{YDEU}. The flavour-violating effects turn out to be quite different for right- and left-handed sfermions. Let us start with the right-handed ones. The light eigenstates turn out to be in the form $f'^c_i = f^c_i -\epsilon\, \alpha^c_i (B-L) F^c$, with $\epsilon\sim 0.06$ and $i=1,2,3$. Therefore, their soft mass matrix is given, in first approximation, by the soft mass matrix $\tilde m^2_{f^c}$ for the three fields $f^c_i$. \Eq{RenLag} shows that the latter have two sizable flavour-dependent interactions, the one with the left-handed messengers, $\lambda_i f^c_i F h$, and the one with the right-handed ones, $\alpha^c_i \phi f^c_i \bar F^c$. In the leading logarithm approximation, the right-handed sfermion mass matrix is given  by
\begin{equation}
\label{eq:llog}
(\tilde m^2_{f^c})_{ij} = m^2_0 \delta_{ij} + \frac{1}{(4\pi)^2} \left[
\left(21 g^2 M^2_{1/2} \delta_{ij} - 12 \lambda_i\lambda_j m^2_0 \right) 
\log\frac{M_c}{M_L} -
\frac{45}{4} \alpha^c_i\alpha^c_j m^2_0 
\log\frac{M_c}{M_R}
\right] 
\end{equation}
at the scale $M_L$ (below that the running is the same as in the MSSM).

In the basis of \eq{YDEU}, in which the fermions are almost diagonal, we have $\lambda_1 = \lambda_2 = 0$, so that the left-handed messengers split the third sfermion mass from the first two but do not induce off-diagonal terms. On the other hand, we expect $\alpha^c_2 \sim \alpha^c_3\neq 0$, so that the interaction with right-handed messengers will give sizable off-diagonal ``2-3'' terms. Numerically, we expect $\delta^{c}_{23} = (\tilde m^2_{f_c})_{23}/m^2_0 = \ord{10^{-1}}$. 

Things are different for the left-handed sfermions. The light eigenstates are essentially $f'_3 = F$, $f'_{1,2} = f_{1,2}$. Therefore, mixed 2-3 terms cannot be generated above the $Z_2$ breaking scale $M_L$.  Hence, we expect the off-diagonal ``2-3'' element of the left-handed sfermion mass matrices to be smaller than in the case of right-handed sfermions. A non-negligible value can be generated through the non-universal A-terms, in turn generated (at the next to leading logarithm level) by the running. 

The large high-scale Yukawa couplings also have an important effect on the diagonal terms of the scalar masses: both the second and the third family masses receive significant contributions. The third family mass would be anyway split by the MSSM running as an effect of the large third family Yukawas. The splitting of the first two families is instead more peculiar and might induce potentially large FCNC and LFV effects in the ``1-2'' sector, once the fermion mass matrices are diagonalized in order to go to the so-called SCKM basis. As a consequence, the pattern of the Yukawa entries involving the first family turns out to be constrained, as we will discuss in Section~\ref{sec:first}. Summarizing, in the basis leading to \eq{YDEU}, the slepton soft masses at $M_L$ are in the form
\begin{equation}
\label{eq:soft}
m^2_{RR} =  \begin{pmatrix}
(m^2_{\tilde{e^c}})_{11} & 0 & 0 \\
0 & (m^2_{\tilde{e^c}})_{22}& (m^2_{\tilde{e^c}})_{23} \\
0 & (m^2_{\tilde{e^c}})_{23} & (m^2_{\tilde{e^c}})_{33}
\end{pmatrix} , \quad
m^2_{LL} =  \begin{pmatrix}
(m^2_{\tilde{l}})_{11} & 0 & 0 \\
0 &  (m^2_{\tilde{l}})_{22} & 0 \\
0 & 0 & m^2_{\tilde{L}}
\end{pmatrix}  .
\end{equation}
The running between $M_c$ and $M_L$ gives $(m^2_{\tilde{e}^c})_{11}>(m^2_{\tilde{e}^c})_{22}>(m^2_{\tilde{e}^c})_{33}$ and
$(m^2_{\tilde{l}})_{11}>(m^2_{\tilde{l}})_{22}>m^2_{\tilde{L}}$. 
The full 1-loop RGEs can be found in the Appendix.

\begin{table}
\begin{center}
\begin{tabular}{|c|c|c|}
\hline
Observable & Bound & Ref.\\
\hline \hline
$\text{BR}(\mu\to e\gamma)$ & $<1.2\times10^{-11}$ & \cite{Brooks:1999pu} \\
$\text{BR}(\tau\to\mu\gamma)$ & $<6.8\times10^{-8}$ & \cite{Aubert:2005ye} \\
$\text{BR}(\tau\to e\gamma)$ & $<1.1\times10^{-7}$ & \cite{Aubert:2005wa}   \\
$\text{BR}(b\to s\gamma)$ & $2.77\times10^{-4}-4.33\times10^{-4}$ (3$\sigma$) & \cite{Barberio:2007cr} \\
$\text{BR}(B_s\to\mu\mu)$ & $<4.7\times10^{-8}$ & \cite{:2007kv} \\
 $(\Delta m_{B_s})^{\rm susy}$ & $< 4.6$ ${\rm ps^{-1}} $ (2$\sigma$)  & \cite{Abulencia:2006ze} \\
 $\delta a^{\rm susy}_\mu$  &  $126\times10^{-11}-478\times10^{-11}$ (2$\sigma$)  & \cite{Passera:2008jk}\\
\hline
\end{tabular}
\end{center}
\caption{\label{Tab:fcnc-constr} Experimental bounds used in the numerical analysis. }
\end{table}

\begin{figure}[t]
\begin{center}
\includegraphics[width=0.95\textwidth]{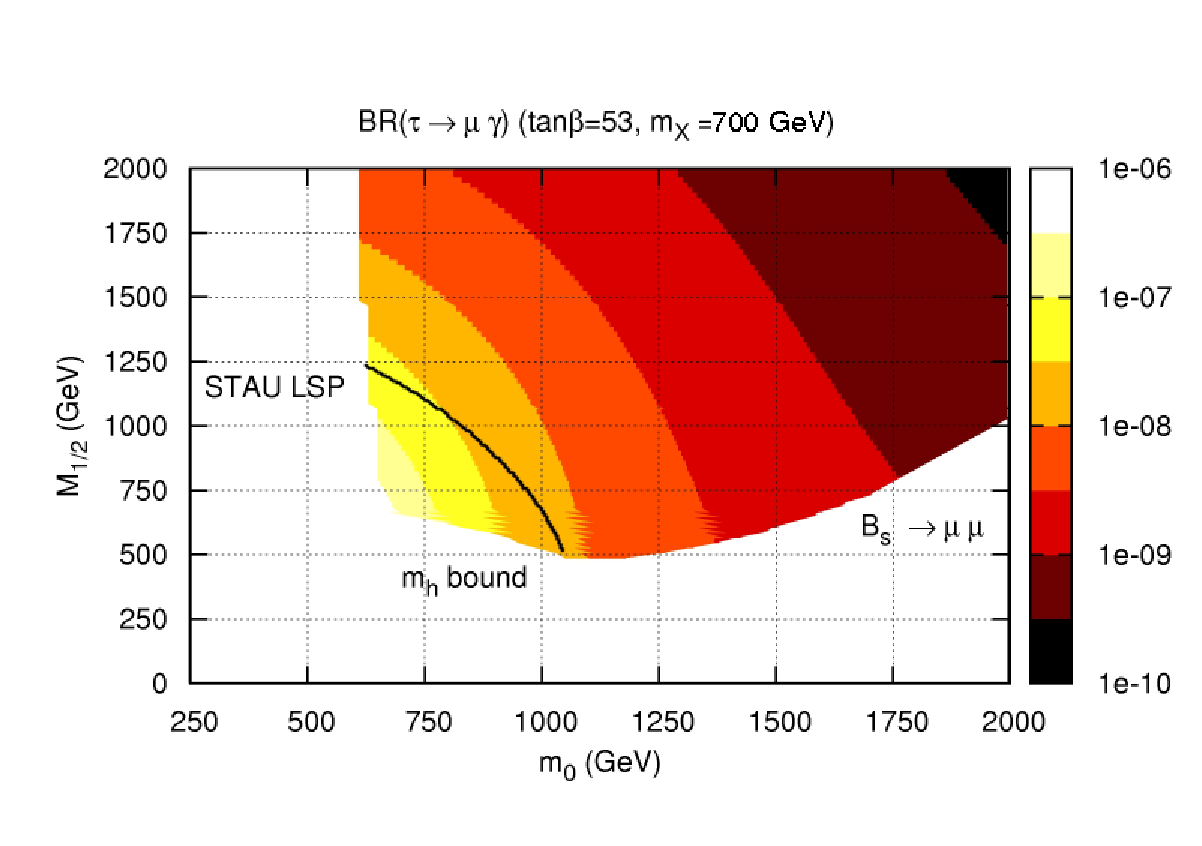}
\end{center}
\caption{Result for $\text{BR}(\tau\to \mu\,\gamma)$ in the ($m_0$, $M_{1/2}$) plane for $\tan\beta\approx 53$, as determined by top-bottom Yukawa unification, $m^0_h = m_0$ and $m_X = 700$ GeV. The black line corresponds to a SUSY contribution to $\delta a^{\rm susy}_\mu = 126\cdot 10^{-11}$. }
\label{fig:tb53}
\end{figure}

\subsection{Numerical analysis}

Exact expressions have been used to compute the branching ratios of LFV decays, $\text{BR}(l_i\to l_j\,\gamma)$~\cite{Hisano:1995cp}, and the SUSY contribution to the anomalous magnetic moment of the muon, $a_\mu=(g-2)_\mu/2$~\cite{Moroi:1995yh}. The SUSY contributions to $B_{d,s}\to \mu\mu$ are estimated by using the formulas in~\cite{Isidori:2002qe}. The branching ratio $\text{BR}(b\to s\,\gamma)$ is computed using the routine {\tt SusyBSG}~\cite{Degrassi:2007kj}. The meson mass splittings ($\Delta m_K$, $\Delta m_D$, $\Delta m_B$, $\Delta m_{B_s}$) are used to constrain the SUSY parameter space. The FCNC processes we have studied and the corresponding experimental measurements or bounds are summarized in Table~\ref{Tab:fcnc-constr}. 

The numerical analysis has been performed by solving the full 1-loop RGEs of the model from the universality scale $M_c$ to $M_L$ and the MSSM RGEs from $M_L$ to $M_{\rm susy} \equiv \sqrt{m_{\tilde t_1} m_{\tilde t_2}}$. The high-scale Yukawas are fixed by requiring a good fit of fermion masses and mixings at $M_Z$. The high-energy parameters which cannot be unambiguously set by low-energy data are chosen to be 1. After the running, the soft mass matrices and $A$-terms are rotated to the SCKM basis, in which the spectrum and the flavour observables are computed.
For each point of the SUSY parameter space, we check that the electroweak symmetry breaking does take place and no tachyonic particles arise. Moreover, to be conservative, we require that the Lightest Supersymmetric Particle (LSP) is the lightest neutralino. Limits on the Higgs and SUSY particles masses from the direct searches at LEP are also imposed. 

We consider two different regimes:
\begin{itemize}
\item Large $\tan\beta$, as determined by top-bottom unification at $M_R$. This is the prediction of the minimal model in Section~\ref{sec:lowE}. We take $m_X = 700\GeV$ in this case (see discussion above). 
\item A lower value of $\tan\beta$, which can be easily obtained in the presence of a mixing in the Higgs sector. We will present the results for the case $\tan\beta=40$, $m_X = 0$.
\end{itemize}
In Fig.~\ref{fig:tb53} we present the prediction for $\text{BR}(\tau\to \mu\,\gamma)$ in the ($m_0$, $M_{1/2}$) plane in the first case. The parameter space is constrained by the requirement of a neutral LSP, by the LEP bound on the Higgs mass, and by $\text{BR}(B_s\to \mu \mu)$.  The latter bound has a significant dependence on the value of $m_X$. In the small $m_X$ limit, in fact, a light CP-odd Higgs enhances the Higgs-mediated contribution\footnote{See \cite{Isidori:2002qe} and references therein.} to $\text{BR}(B_s\to \mu \mu)$, especially in the large $m_0$, moderate $M_{1/2}$ region. Larger values of $m_X$ increase $m_A$, thus alleviating the bound from $B_s\to \mu\mu$. 

\begin{figure}[t]
\begin{center}
\includegraphics[angle=-90,width=0.85\textwidth]{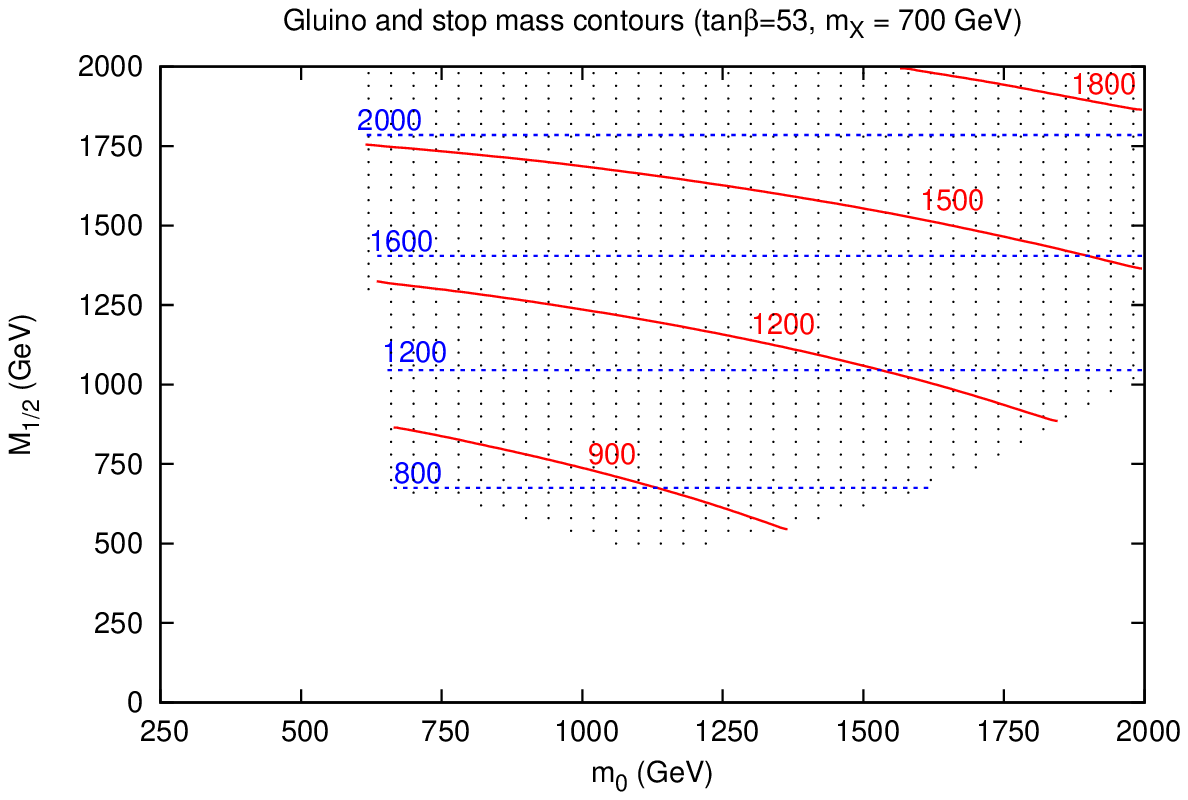}
\end{center}
\caption{Contour plot for the gluino (blue, dashed lines) and the lightest stop (red, solid lines) masses, for the same choice of parameters as in Fig.~\ref{fig:tb53}.}
\label{fig:tb53masses}
\end{figure}

The present limit $\text{BR}(\tau\to\mu\,\gamma)\lesssim 6.8\cdot 10^{-8}$ only gives a weak constraint on the parameter space, $(m_0,\, M_{1/2})\lesssim 1 \,\,{\rm TeV}$. A Super $B$-Factory able to reach a sensitivity of $10^{-9}$ \cite{Bona:2007qt} would test a large portion of the parameter space shown. The black line in Fig.~\ref{fig:tb53} corresponds to $\delta a^{\rm susy}_\mu = 126\cdot 10^{-11}$. In the region below the black line, the supersymmetric contribution to the muon magnetic moment would reduce the present tension\cite{Passera:2008jk} between the experimental measurement of the muon magnetic moment and the SM prediction below the $2\sigma$ level. 

In Fig.~\ref{fig:tb53masses} we show contour plots of the lightest stop and gluino masses, for the same choice of the parameters in Fig.~\ref{fig:tb53}. Comparing Fig.~\ref{fig:tb53masses} with Fig.~\ref{fig:tb53}, we see that, for the case considered here, a Super $B$-factory bound $\text{BR}(\tau\to\mu\,\gamma)\lesssim 10^{-9}$ would be able to test almost completely the parameter space for SUSY masses within the LHC reach.

\begin{figure}[t]
\begin{center}
\includegraphics[width=0.95\textwidth]{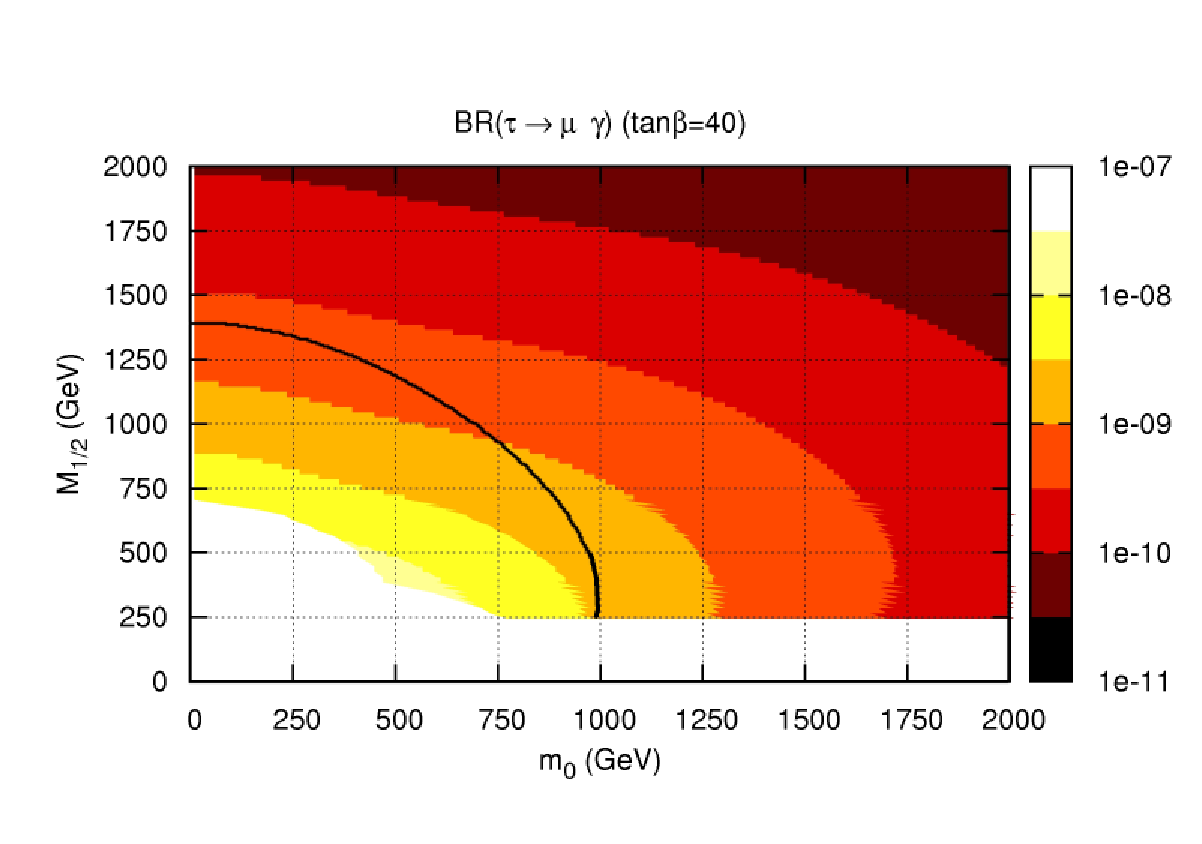}
\end{center}
\caption{Same as Fig.~\ref{fig:tb53} for the case with $\tan\beta=40$, $m^0_h = m_0$ and $m_X = 0$.}
\label{fig:tb40}
\end{figure}

In Fig.~\ref{fig:tb40} the results for the $\tan\beta=40$, $m_X=0$ case are shown for $m_h^0=m_0$. In this case the parameter space is much less constrained. The low $M_{1/2}$ region is excluded by LEP direct searches for SUSY particles, while $b\to s\,\gamma$ and $B_s\to \mu\mu$ constrain the low $m_0$, low $M_{1/2}$ regime. The present limit on $\text{BR}(\tau\to\mu\,\gamma)$ does not give rise to additional constraints and the Super $B$-factory sensitivity should test the parameter space up to $(m_0,\, M_{1/2})\sim 1\, {\rm TeV}$. The $(g-2)_\mu$ result favours in this case a region almost completely within the reach of the Super $B$-factory. 

\begin{figure}[t]
\begin{center}
\includegraphics[angle=-90,width=0.85\textwidth]{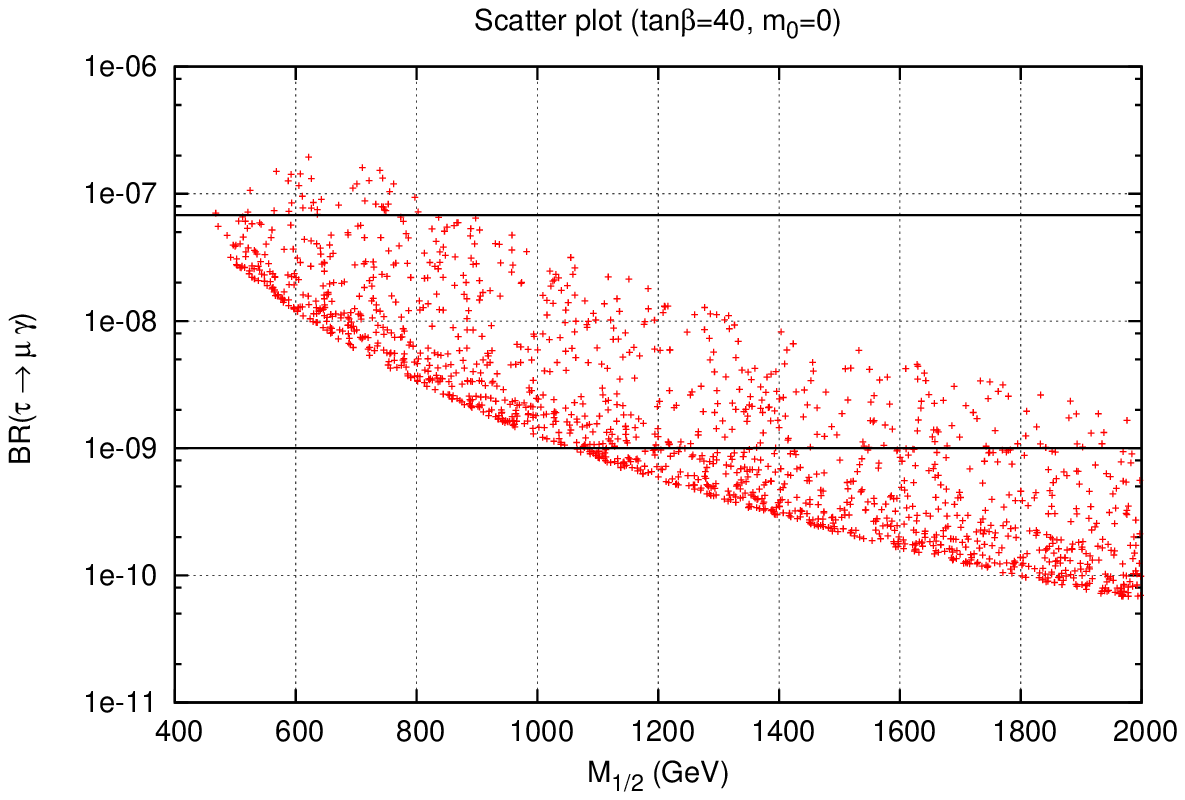}
\end{center}
\caption{$\text{BR}(\tau\to\mu\,\gamma)$ vs $M_{1/2}$ for $\tan\beta=40$ and ``gaugino-mediation''-inspired boundary conditions ($m_0=m_h^0=m_X=0$). The unknown $\ord{1}$ couplings are randomly varied between 0.5 and 1.5. Horizontal lines represent the present bound ($6.8\times 10^{-8}$) and the future limit ($10^{-9}$) on $\text{BR}(\tau\to\mu\,\gamma)$.  }
\label{fig:tb40-scatter}
\end{figure}

As previously mentioned, some of the high-energy Yukawas are not determined by the fit of the fermion masses and mixings and have been fixed to $\ord{1}$ values. In order to show the impact of those $\ord{1}$ parameters on $\text{BR}(\tau\to\mu\,\gamma)$, we display in Fig.~\ref{fig:tb40-scatter} a scatter plot where all the unknown Yukawas are independently varied between 0.5 and 1.5. We consider in this case ``gaugino-mediation'' boundary conditions with $m_0=m_X=0$. The figure shows that the $\ord{1}$ parameters can conspire to enhance $\text{BR}(\tau\to\mu\,\gamma)$ by up to one order of magnitude.

The WMAP constrain on the dark matter relic density~\cite{Spergel:2006hy} can also be accounted for by a neutralino thermal relic within this model. Two examples of parameter space points leading to the correct density are exhibited in Tab.~\ref{Tab:points1}. Point {\bf A} is an example of the unified top-bottom Yukawa regime, point {\bf B} is an example of the  $\tan\beta=40$ case, with gaugino-mediation boundary conditions. For point {\bf A}, the WMAP bound on cold dark matter is satisfied by stau-neutralino co-annihilation~\cite{Ellis:1998kh}. In the case of point {\bf B} the correct relic density is given by the LSP annihilation through the CP-odd Higgs s-channel exchange~\cite{Drees:1992am}. In Tab.~\ref{Tab:points2} we show the corresponding SUSY and Higgs spectrum, the predictions for the $b\to s\,\gamma$ branching ratio, the SUSY contribution to the magnetic moment of the muon $\delta a^{\rm susy}_\mu$, $\tau\to \mu\,\gamma$ branching ratio and the DM relic density $\Omega_{\rm DM}\,h^2$, which has been computed using the routine {\tt micrOMEGAs}~\cite{Belanger:2006is}. 
For both points LHC should observe several SUSY particles, even if the spectrum of point {\bf A} is heavier. Both points give a  $\text{BR}(\tau\to \mu\,\gamma)$  within the sensitivity of the proposed Super $B$-factory, while point {\bf B} gives a better agreement with $(g-2)_\mu$. 

\begin{table}[t]
\begin{center}
\begin{tabular}{|c||c|c|c|c|c|}
\hline
 Point & $\tan\beta$ & $m_0$ & $m_h^0$ & $m_X$ & $M_{1/2}$ \\
\hline 
{\bf A} & 53 & 650 & 650 & 700 & 1400 \\
{\bf B} & 40 &  0 &  0 &  0 & 850 \\
\hline
\end{tabular}
\end{center}
\caption{Two points of the parameter space accounting for the dark matter relic density. The masses are expressed in GeV.\label{Tab:points1}  }
\end{table}

\begin{table}[t]
\begin{center}
\begin{small}
\begin{tabular}{|c||c|c|c|c|c|c|c|c|c|c|}
\hline
 & $m_{\tilde{t}_1}$ & $m_{\tilde{g}}$ & $m_{\tilde{\tau}_1}$ & $m_{\tilde{\chi}^0_1}$ & $m_h$ & $m_A$ & $\text{BR}(b\to s\,\gamma)$ &
 $\delta a^{\rm susy}_\mu$ & $\text{BR}(\tau\to \mu\,\gamma)$ & $\Omega_{\rm DM}\,h^2$  \\
\hline 
{\bf A} & 1291 & 1661 & 360 & 339 & 119 & 1000 & $3.15\times 10^{-3}$  & $105\times 10^{-11}$ & 
  $2.4\times 10^{-8}$ & 0.095 \\
{\bf B} & 723 & 1023  & 243  &  202 & 117 & 353 & $3.16\times 10^{-3}$ & $316\times 10^{-11}$ & $4\times 10^{-9}$ &  0.106 \\
\hline
\end{tabular}
\end{small}
\end{center}
\caption{Spectrum and predictions for the two parameter space points in Table~\ref{Tab:points1}. The masses are expressed in GeV. \label{Tab:points2}  }
\end{table}

Finally, let us comment on possible LFV effects at the LHC. In the moderate $\tan\beta$
case, the structure of the radiative contributions in \eq{llog} is particularly interesting in this respect: $\tilde m^2_{ij} = m^2_0 \delta_{ij} + \sigma c_{ij} m^2_0$, where $c_{ij}$ is a matrix with $\ord{1}$ entries and $\sigma$ accounts for all the rest, including the loop factors (in the large $\tan\beta$ case, the bottom Yukawa radiative contributions to $c_{33}$ dominate the matrix $c_{ij}$ and the mixing angle typically turns out to be
too small to give rise to measurable effects). The interesting feature is that this structure gives rise to large mixing between the second and third family independently of how small $\sigma$ is. This might give large effects in $\tau\to\mu\gamma$ transitions. However, by making $\sigma$ small these effects can be kept under control. Still, the LFV effects at colliders do not get suppressed, as long as $|\tilde m_2 - \tilde m_3| \gtrsim  \Gamma$\cite{ArkaniHamed:1996au}, where $\Gamma$ is the slepton width, because the mixing angle remains large in the $\sigma\to 0$ limit. Note that collider and low energy effects ($\text{BR}(\tau\to\mu\gamma)$)are complementary, as the former are interesting when $\tan\beta$ is not too large, while the latter are enhanced in the large $\tan\beta$ regime.

Let us summarize this Section. We studied the predictions of the flavor model of Section~\ref{sec:lowE} for FCNC and LFV effects. For this, we assumed universal soft SUSY-breaking terms at a high scale $M_c > M_R$, which will be identified with the compactification scale of the full 5D model. We studied the RG evolution of the parameters to low energy for two different regimes of $\tan\beta$. The comparison with low-energy data (fermion masses and mixings) gives constraints on some of the Superpotential couplings, which are of $\ord{1}$ as expected. All couplings which are not fixed by data are assumed to be $\ord{1}$. We find the following results:
\begin{itemize}
\item in the large $\tan \beta$ regime ($\tan \beta = 53$), the present limit on $BR(\tau \to \mu \gamma)$ gives only weak constraints on the SUSY parameter space, $(m_0,\, M_{1/2})\sim 1\, {\rm TeV}$, while the expected sensitivity of a Super $B$-Factory could probe almost completely the parameter space for SUSY masses within the LHC reach;
\item for a lower value of $\tan \beta = 40$, the parameter space is presently unconstrained and will be tested by future experiments up to $(m_0,\, M_{1/2})\sim 1\, {\rm TeV}$.
\end{itemize}

\section{An SO(10) embedding in 5 dimensions}
\label{sec:SO(10)}

We now embed the PS model illustrated in Section~\ref{sec:lowE} in a five-dimensional, supersymmetric, SO(10) model. What follows is not the only possible embedding in a 5D unified model and as such it should be considered as a proof of existence of a unified version of the model. 

The benefits of considering unification in the presence of one or more extra-dimension with an inverse size of the order of the Grand Unification Theory (GUT) scale are well known and include an easy implementation of doublet-triplet splitting and the suppression of dimension five operators contribution to proton decay \cite{Hall:2001pg,SU5OFGUTS}. In our case, the presence of the fifth dimension is welcome also because it provides more freedom on the light field spectrum, which in turn helps maintaining (and in some case improving) the successful $\alpha_s$ prediction in the MSSM. The use of SO(10) is dictated by the need of having PS as a subgroup. Earlier work on SO(10) in five dimensions can be found in \cite{SO10OFGUTS}. 

The fifth dimension is compactified on a $S^1/(Z_2\times Z'_2)$ orbifold. SO(10) is broken to $G_\text{PS}$ on the $y=\pi R/2$ brane and unbroken on the $y=0$ brane ($R$ is the compactification radius, $y$ the coordinate on the fifth dimension, $y\sim y+2\pi R$, $y\sim -y$, $y'\sim -y'$, $y' = y+\pi R/2$). The 4D $N=2$ supersymmetry is broken to $N=1$ supersymmetry on both the four-dimensional branes. The $Z_2$ parities relate the values of the fields as follows: $\Phi(x,y) = P \Phi(x,-y)$,  $\Phi(x,y^\prime) = P^\prime \Phi(x,-y^\prime)$, where $P^2 = P'^2 = 1$. In an appropriate basis, each field $\Phi$ can be classified by its eigenvalues $(\pm 1, \pm 1)$ under $(P,P')$. The expansion in Kaluza-Klein (KK) modes is then
\globallabel{eq:KKexpansion}
\begin{align}
\Phi_{++}(x,y) &= \sqrt{\frac{4}{\pi R}} \sum_{n=0}^{\infty} \frac{1}{(\sqrt{2})^{\delta_{n,0}}} \Phi_{++}^{(2n)}(x) \cos{\frac{2ny}{R}} \mytag \\
\Phi_{+-}(x,y) &= \sqrt{\frac{4}{\pi R}} \sum_{n=0}^{\infty} \Phi_{+-}^{(2n+1)}(x) \cos{\frac{(2n+1)y}{R}}\mytag\\
\Phi_{-+}(x,y) &= \sqrt{\frac{4}{\pi R}} \sum_{n=0}^{\infty} \Phi_{-+}^{(2n+1)}(x) \sin{\frac{(2n+1)y}{R}}\mytag\\
\Phi_{--}(x,y) &= \sqrt{\frac{4}{\pi R}} \sum_{n=0}^{\infty} \Phi_{--}^{(2n+2)}(x) \sin{\frac{(2n+2)y}{R}}. \mytag
\end{align}
As usual, only $\Phi_{++}$ and $\Phi_{+-}$ ($\Phi_{++}$ and $\Phi_{-+}$) are possibly non-vanishing at the $y=0$ ($y=\pi R/2$) brane and only $\Phi_{++}$ has a massless zero mode. 

The supersymmetric structure can be formulated in terms of the superfield language of the unbroken four-dimensional $N=1$ supersymmetry. The 5D vector multiplet consists of a vector and a chiral multiplet, $V$ and $\Phi$, while the 5D hypermultiplet decomposes into two chiral multiplets in conjugate representations, say $H$ and $\tilde{H}$. The orbifold parities of the vector multiplet components can be chosen in such a way that $V = V^{\text{PS}}_{++} + V^{\text{SO(10)/PS}}_{+-}$, $\Phi = \Phi^{\text{PS}}_{--} + \Phi^{\text{SO(10)/PS}}_{-+}$. SO(10) is thus unbroken at $y=0$ (the ``SO(10) brane'') and broken to PS at $y= \pi R/2$ (the ``PS brane''), and $N=2$ supersymmetry is broken to $N=1$ on both branes, as anticipated. Let us now consider a bulk SO(10) hypermultiplet $(H,\tilde H)$. In most cases $H$ and $\tilde H$ split into two PS components $H = H_1 + H_2$,  $\tilde H = \tilde H_1 + \tilde H_2$. Let $H_1$ be the ``++'' mode, $H_1 = (H_1)_{++}$. The relative orbifold parities are then dictated by the invariance of the bulk action: $H_2 = (H_2)_{+-}$, $\tilde H_1 = (\tilde H_1)_{--}$, $\tilde H_2 = (\tilde H_2)_{-+}$. All the degrees of freedom of the vector multiplet but the PS gauge bosons and gauginos get mass at the compactification scale $M_c \equiv 1/R$ or higher. Some of the fields in the effective model of Section~\ref{sec:lowE} are embedded in bulk fields. In this case, they correspond to the ``++'' (zero) modes of those fields. All the other degrees of freedom of those bulk fields get a mass at the scale $1/R$ or higher. The fields heavier than the compactification scale can be ignored in first approximation. They will be discussed in the context of gauge coupling unification. 
\medskip

\begin{table}[htdp]
\begin{equation*}
\begin{array}{|c|ccc|ccccccc|cccc|}
\hline
& \psi_i & \psi' & \bar\psi' & F & F_c & \bar F & \bar F_c & h & \phi & S_j & F_c' & \bar F'_c & X_c & \Sigma \\ \hline
\text{Localization} & \multicolumn{3}{c|}{\text{SO(10)}} & \multicolumn{7}{c|}{\text{bulk}} & \multicolumn{4}{c|}{\text{PS}} \\
\text{Gauge repr} & 16 & 16 & \overline{16} & 16 & 16 & \overline{16} & \overline{16} & 10 & 45 & 1 & (1,2,\bar 4) & (1,2,4) & (1,3,1) & (1,1,15) \\
\text{U(1)}_R & 1 & 0 & 0 & 1 & 1 & 1 & 1 & 0 & 0 & 1 & 0 & 0 & 0 & 1 \\
Z_{24} &  -7 & 9 & 1 & -6 & -6 & -6 & -6 & -11 & -11 & 6 & 2 & 2 & 12 & 4 \\
\hline
\end{array}
\end{equation*}
\caption{Embedding of the fields in Table~\ref{tab:fields} in the 5D SO(10) model ($i,j=1,2,3$), localization in the fifth dimension and quantum numbers under the relevant symmetries.}
\label{tab:flavourfields}
\end{table}

Let us now come to the embedding of the effective model of Section~\ref{sec:lowE}. The PS vector fields are identified with the zero modes of $V^\text{PS}_{++}$. The fields in Table~\ref{tab:fields} are embedded in the fields in Table~\ref{tab:flavourfields}. The fields $f_i$ and $f^c_i$, $i=1,2,3$, making up the three MSSM families in the unbroken $\mathbf{Z}_2$ limit, become the three $\psi_i$ on the SO(10) brane. The successful SO(10) predictions of the SM fermion gauge quantum numbers is therefore maintained. The Higgs $h$, the $\mathbf{Z}_2$ breaking field $\phi$, and the messengers $F$, $F_c$, $\bar F$, $\bar F_c$ become the zero modes of the ++ component of the corresponding bulk fields (with an abuse of notation we denote the bulk field by the symbol that would be used for its zero mode component). The PS-breaking fields $F'_c$, $\bar F'_c$, $X_c$, and the SU(2)$_R$-breaking messenger field $\Sigma$ live on the PS brane. The spectrum in Table~\ref{tab:flavourfields} also includes 3 bulk singlets $S_j$, $j=1,2,3$, and the $\psi'+ \bar \psi'$ on the SO(10) brane. The latter fields play a role in the neutrino sector. 

We aim at exhibiting a full model taking care of the vevs used in the generation of the flavour structure, among the other things. The fields involved in those (implementation-dependent) ``side'' aspects of the model are listed in Table~\ref{tab:additionalfields}. Some of them are needed, as mentioned, to generate the necessary vevs (the singlets, essentially), some to get a field content able to preserve an MSSM 1-loop unification all the way up to the unification scale~\cite{CFRZ}  ($H_6$, $x_c$, $x$, $\Omega$), some to avoid unwanted Goldstones and to set each field at the appropriate scale. 

\begin{table}[htdp]
\begin{equation*}
\begin{array}{|c|cccc|ccc|ccccc|}
\hline
& \Phi & Y_{10} & Y'_{10} & H & H_6 & \theta^\pm & \Theta^\pm & Y_\text{PS} & Y'_\text{PS} & x_c & x & \Omega \\ \hline
\text{Localization} & \multicolumn{4}{c|}{\text{SO(10)}} & \multicolumn{3}{c|}{\text{bulk}} & \multicolumn{5}{c|}{\text{PS}} \\
\text{Gauge repr} & 45 & 1 & 1 & 10 & 10 & 1 & 1 & 1 & 1 & (1,3,1) & (3,1,1) & (2,2,6) \\
\text{U(1)}_R & 2 & 2 & 2 & 2 & 2 & 0 & 0 & 2 & 2 & 2 & 1 & 0\\
Z_{24} & -10 & -10 & 0 & 2 & 1 & \pm 3 & \mp 2 & -4 & 0 & 3 & 3 & -9 \\
\hline
\end{array}
\end{equation*}
\caption{Additional fields involved in different aspects of the 5D SO(10) model}
\label{tab:additionalfields}
\end{table}

Tables~\ref{tab:flavourfields} and~\ref{tab:additionalfields} show the U(1)$_R$ assignment of the fields. The U(1)$_R$ symmetry we are considering is not directly related to the bulk SU(2)$_R$ symmetry characterizing 5D supersymmetry. The latter is broken to a U(1)$'_R$ by boundary conditions and can easily be strongly broken spontaneously. The former $R$-symmetry contains the $R$-parity symmetry $R_P$ used in Section~\ref{sec:lowE}~\footnote{The U(1)$_R$ could for example be broken together with supersymmetry down to an $R$-parity if the superfield breaking supersymmetry has $R=0$.}. It will play a role in suppressing proton decay and naturally explains why the MSSM Higgs fields stay light \cite{Hall:2001pg}. The discrete $Z_{24}$ symmetry is used to constrain the superpotential and contains the $\mathbf{Z}_2$ of Section~\ref{sec:lowE}, to which it is spontaneously broken above the scale $M_R$ ($\mathbf{Z}_2$ is broken as needed to reproduce the results in Section~\ref{sec:lowE}). 

\subsection{The strong coupling order parameter}

Two nice features of the model we are proposing are the possibility to relate the $\ord{1}$ parameters to strong couplings and the fact that some of the order parameters (all except two, as we will see) are identified with hierarchies among strong couplings involving different types of fields. In order to show this, we first introduce properly normalized, dimensionless fields. 

We assume that the theory approaches a strongly interacting regime at the cutoff scale $\Lambda$. Naive dimensional analysis (NDA) suggests to write the action in terms of normalized derivatives $\hat\partial = \partial/\Lambda$ and of dimensionless chiral and vector superfields $\hat\phi$, $\hat V$, related to the canonically normalized fields $\phi$, $V$ by 
\begin{equation}
\label{eq:naturalunits}
\phi_4 = \hat\phi_4 \fracwithdelims{(}{)}{\Lambda^2}{l_4}^{1/2}, \quad 
\phi_5 = \hat\phi_5 \fracwithdelims{(}{)}{\Lambda^3}{l_5}^{1/2}, \quad 
V_4 = \hat V_4 \fracwithdelims{(}{)}{\Lambda^2}{l^V_4}^{1/2}, \quad 
V_5 = \hat V_5 \fracwithdelims{(}{)}{\Lambda^3}{l^V_5}^{1/2},
\end{equation} 
where the index 4 (5) denotes brane (bulk) fields. When expressed in terms of the dimensionless fields above, the brane superpotential acquires the form 
\begin{equation}
\label{eq:braneW}
W_\text{brane}(\phi_i) = \frac{\Lambda^3}{l_4} \hat W(\hat\phi_i),
\end{equation} 
where $\hat W$ does not contain dimensionful parameters and its expansion is expected to involve $\ord{1}$ coefficients \cite{Chacko:1999hg}\footnote{Note that the derivation in \cite{Chacko:1999hg} assumes an infinite extra-dimension. See\cite{Hall:2002ci} for a more detailed approach.}. 

The values of the dimensionless coefficients $l^{(V)}_{4,5}$ leading to $\ord{1}$ coefficients in $\hat W$ (defined of course themselves up to $\ord{1}$ factors) depend on the theory under consideration and may be different for different fields. The guideline provided by NDA is that $l_D$ is just the loop factor in $D$ dimensions: $l_D= (4\pi)^{D/2} \Gamma(D/2)$. In our case, we will use the same factor $l_4$ ($l_5$) for all the chiral brane (bulk) superfields (superpotential couplings), while we keep the possibility of having a different normalization for the vector fields (gauge couplings). This is because the gauge couplings are qualitatively different in that the coefficients of the gauge loop expansion grow with the number of charged matter fields. With the field content in the Tables~\ref{tab:flavourfields} and~\ref{tab:additionalfields}, we expect $l_V$ to be smaller by a factor $\ord{5}$~\cite{Hall:2002ci}. 

Whatever are the precise values of the coefficients $l^{(V)}_{4,5}$, the important point for our purposes is that they lead to small hierarchies that contribute to account for the fermion mass hierarchies \cite{Hall:2001rz} . In practice, the relevant order parameter turns out to be 
\begin{equation}
\lambda \equiv 1/l_4^{1/4} \approx 0.24 \approx \sqrt{\epsilon}. 
\end{equation}
The numerical value is compatible with the NDA prediction for $l_4$. However, it is actually chosen in order to be able to best fit the numerical values in the following. The reason why this is the relevant parameter is that it enters the expected values of mass terms, as we now see. When written in terms of the dimensionless fields, the mass terms in $\hat W$ will be dimensionless numbers, say $\eta$. In the strong coupling regime, we expect $\eta = \ord{1}$, but smaller values are of course allowed. In terms of canonically normalized fields, the mass term shows a dependence on $\lambda$: 
\begin{equation}
\label{eq:mass}
M \sim \eta\, \lambda^{n_B} \Lambda ,
\end{equation}
where $n_B$ is the number of bulk fields involved in the mass term ($n_B = 0,1,2$). 
In order to understand the above formula, one should first note that a term in $\hat W$ involving $n_B$ bulk fields and $n_b$ brane fields will give rise in this strong coupling regime to an effective 4D  coupling $g$ of order
\begin{equation}
\label{eq:couplings}
g\sim \fracwithdelims{(}{)}{2l_5}{\pi R\Lambda}^{n_B/2} l_4^{n_b/2-1}
\end{equation}
(we neglect the running, which can be significant). The effective model in Section~\ref{sec:lowE} assumes the couplings to be $\ord{1}$. Because it will turn out that those couplings come from interactions involving $n_B=2$ bulk and $n_b = 1$ brane fields, we can write $2l_5/(\pi R\Lambda)$ as $l_4^{1/2} = 1/\lambda^{2}$, up to $\ord{1}$ factors. This is why only the $\lambda$ parameter enters \eq{mass}. Also, this allows to relate $\Lambda R$ to $l_{4,5}$. Assuming that $l_5$ is such that $2l_5/\pi \sim 100$, we have $\Lambda R \sim 5$. We can check the consistency of the numbers estimating the size of the gauge couplings at the cutoff scale. This is determined by $l_5/l^V_5 \sim 5$, which gives $g^2_{4D}(\Lambda) \sim l^V_5/(\lambda^2 l_5)\sim 3.5$, which is close to the (radiatively enhanced) value of $g^2_{4D}$ we find in Section~\ref{sec:unification}. 

\subsection{Brane superpotentials}

The (normalized) superpotentials on the SO(10) and PS branes are
\begin{equation}
\label{eq:Wpieces}
\hat W_\text{SO(10),PS} = \hat W_\text{SO(10),PS}^\text{flav} + \hat W_\text{SO(10),PS}^\text{vevs} + \hat W_\text{SO(10),PS}^\text{mass} .
\end{equation} 
The $\hat W^\text{flav}_\text{SO(10),PS}$ parts are directly related to the superpotential in \eq{RenLag} and therefore to the SM flavour structure. They are (keeping only ``++'' components of bulk fields)
\globallabel{eq:Wflavour}
\begin{align}
\hat W^\text{flav}_\text{SO(10)} &= 
\lambda_i\hat\psi_i \hat F \hat h + \lambda^c_i \hat\psi_i \hat F_c \hat h + 
\alpha_i \hat\psi_i \hat{\bar F}\hat \phi + 
\alpha^c_i \hat{\psi_i} \hat{\bar F}_c \hat\phi + a_{ij} \hat{\bar\psi}' \hat S_i \hat\psi_j \mytag, \\
\hat W^\text{flav}_\text{PS} &= 
a\hat {\bar F}_c \hat X_c \hat F_c + \bar\sigma_c\hat{\bar F}'_c \hat\Sigma \hat F_c + \sigma_c \hat{\bar F}_c \hat\Sigma \hat F'_c + b\frac{\hat F'_c \hat X_c \hat F'_c}{2} \hat\Sigma^2 + b_i \hat{\bar F}'_c
 \hat S_i
 \hat F_c \hat\Theta_+ +c_i \hat{\bar F}_c
 \hat S_i
 \hat F'_c \hat\Theta_+ .  \mytag
\end{align} 
The last terms affect the singlet neutrino mass matrix. All couplings are $\ord{1}$, as predicted by the strong coupling assumption. Despite the same notation, they are not exactly the same as the corresponding parameters in \eq{RenLag}, but the difference is an irrelevant $\ord{1}$ factor.

Some of the $R=0$ fields in the superpotentials above get a vev due to (here and below we omit $\ord{1}$ coefficients)
\globallabel{eq:Wvevs}
\begin{align}
\hat W^\text{vevs}_\text{SO(10)} &= 
\hat Y_{10} (\hat {\bar\psi}'\psi' - \hat \theta^2_+ \hat\Theta^2_-) + 
\hat Y'_{10} (\hat \theta_+ \hat \theta_- -\epsilon^2_{10}) + 
\hat {\bar \psi}' \hat \Phi\, \hat \psi' +
\hat \theta_- \hat \phi \,\hat \Phi
\mytag, \\
\hat W^\text{vevs}_\text{PS} &= 
\hat Y_\text{PS} (\hat {\bar F}'_c \hat F'_c - \hat \Theta^2_-) + 
\hat Y'_\text{PS} (\hat \Theta_+ \hat \Theta_- -\epsilon^2_\text{PS}) +
\hat \theta_- \hat \Theta_+^2 \hat {\bar F}'_c\, \hat x_c \hat F'_c +
\hat \theta_+ \hat \Theta_-^3 \hat x_c\, \hat X_c. 
\mytag
\end{align} 
The superpotentials above contain two anomalously small coefficients $\epsilon_{\text{PS}}\sim\lambda$ and $\epsilon_{10}\sim\lambda^2$ characterizing the PS and SO(10) branes respectively. Together with the dynamically generated $\lambda$, $\epsilon_{\text{PS}}$ and $\epsilon_{10}$ are the seeds of the fermion hierarchies. It is tempting to generate dynamically $\epsilon_{\text{PS}}$ and $\epsilon_{10}$ in terms of $\lambda$, but this is beyond the scope of this paper. Another aspect of the model that is left to further investigation is the suppression of the mass term arising from the vev of the SU(2)$_R$-triplet component of $\phi$ in \eq{Wflavour}. This suppression is necessary in order to preserve the $m_\mu/m_s$ ratio obtained in Section~\ref{sec:lowE}. The $F$-term equations give 
\begin{equation}
\label{eq:seeds}
\VeV{\big}{\hat \Theta_\pm} \sim \VeV{\big}{\hat F'_c} \sim \VeV{\big}{\hat {\bar F}'_c} \sim \VeV{\big}{\hat X_c} \sim \lambda, \qquad
\VeV{\big}{\hat \theta_\pm}  \sim \lambda^2, \qquad \VeV{\big}{\hat \psi'} \sim \VeV{\big}{\hat {\bar \psi}'} \sim \lambda^3, \qquad
\VeV{\big}{\hat \phi} \sim \lambda^4 .
\end{equation} 

Finally, we need to provide mass terms for some otherwise light fields. These are provided by
\begin{equation}
\label{eq:neededmasses}
\hat W^\text{mass}_\text{PS} = 
\hat \theta_- \hat \Theta_+ \hat {\bar F}'_c \hat {\bar F}'_c \hat H_6 +
\hat \theta_- \hat \Theta_+ \hat F_c'  \hat F_c' \hat H_6 + \frac{\hat \Theta_+^3}{2} \hat x^2
+\hat \psi'\hat \psi'\hat H\hat \Theta^2_- + \hat{\bar\psi}'\hat{\bar\psi}'\hat H\hat \Theta^2_+  .
\end{equation} 
Also relevant are two operators involving fields with no zero mode. The bulk field we denoted by $\phi$ is an hypermultiplet corresponding to 4 PS bulk fields with different orbifold parities: $(\phi_\text{PS})_{++}$, the field whose zero mode enters \eq{RenLag}, $(\phi_\text{226})_{+-}$, its SO(10) complement, and the conjugated fields $(\tilde\phi_\text{PS})_{--}$ and $(\tilde \phi_{226})_{-+}$. The operators $(\phi_\text{226})_{+-}$ and $(\tilde\phi_{226})_{-+}$ have no zero mode but they are relevant for our purposes. They appear in fact in the following two operators: $\hat \Omega \hat \Theta_+(\hat {\tilde \phi}_{226})_{-+} $, on the PS brane, and $\hat \Phi_{226} \hat \theta_-(\hat {\phi}_\text{226})_{+-} $, on the SO(10) brane. When integrating out the heavy $({\tilde \phi}_{226})_{-+}$, $({\phi}_\text{226})_{+-}$ fields, one obtains the operator $\hat \Theta_+ \hat \theta_- \hat \Omega \hat \Phi_{226} $, involving fields from two different branes\footnote{Let $b^{10}_i$, $i=1\ldots h$ be SO(10) brane fields, $b^\text{PS}_j$, $j=1\ldots k$ be PS brane fields and $B$ be a bulk field. After integrating out the heavy bulk fields, the operators $a_{10} \hat b^{10}_1\ldots \hat b^{10}_h \hat B_{(+,-)}$ and $a_\text{PS} \hat b^\text{PS}_1\ldots  \hat b^\text{PS}_k \hat {\tilde B}_{(-+)}$ on the two branes give rise to the effective interaction $a_{10} a_\text{PS} \hat b^{10}_1\ldots \hat b^{10}_h \hat b^\text{PS}_1\ldots  \hat b^\text{PS}_k$.\label{foo:branebrane}}. The latter gives a mass term at the scale $M_R$ coupling the fields $\Omega$ and $\Phi_{226}$.

\subsection{Scales, spectrum, and unification}

We now illustrate the spectrum we obtain for the heavy fields and in particular we relate the scales $M_R$ and $M_L$  of the right- and left-handed messengers defined in Section~\ref{sec:lowE} to the cutoff $\Lambda$. In order to do that, we make an extensive use of \eq{mass}. 

We take $\Lambda \approx 10^{17}\GeV$. The right-handed messengers in $F_c$, $\bar F_c$, as well as $\Sigma$, get a mass 
\begin{equation}
\label{eq:MR}
\ord{\lambda^3 \Lambda} \sim 2\cdot 10^{15}\GeV \equiv M_R .
\end{equation}
The up quark sector involves a mixed mass term arising from the vevs of $F'_c$, $\bar F'_c$, which is enhanced by a factor 1/$\lambda$, $V_c \sim M_R/\lambda \approx M_R/\sqrt{\epsilon}$. As discussed in Section~\ref{sec:lowE}, such an enhancement accounts for the smallness of $m_c/m_t$ (and will give a threshold correction to gauge coupling unification). The masses $M_R$ and $V_c$ are different despite they arise from dimensionless vevs of the same order because the corresponding mass terms contain a different number of bulk fields, in agreement with \eq{mass}. 

Once $\mathbf{Z}_2$ is broken by the vev of $\phi$, the messengers and the would be light families get mixed by a mass term
\begin{equation}
\label{eq:ML}
\ord{\lambda^5 \Lambda} \sim 3 \cdot 10^{14}\GeV \equiv M_L ,
\end{equation}
so that $\epsilon = M_L/M_R \approx \lambda^2\approx 0.06$. The two mass terms mixing the singlets $S_i$ with $N_c$ and $n^c_i$ are both $\ord{\lambda M_R}$, half way between $M_L$ and $M_R$. 

Below the compactification scale $M_c = 1/R$, SO(10) is broken to PS and the spectrum is made of the PS vector fields, the brane fields, and the zero modes of the bulk fields in the Tables~\ref{tab:flavourfields} and~\ref{tab:additionalfields}. It is interesting (although in part cooked up) that this constitutes a retarding PS ``magic'' field content, with the terminology of~\cite{CFRZ}. This means that such a field content would exactly preserve (but delay to a higher GUT scale) a MSSM 1-loop gauge coupling unification, despite the field content does not correspond to a complete SU(5) representation. A systematic study of such ``magic'' field contents can be found in~\cite{CFRZ}. We will make extensive use of such contents in the following. 

Let us now consider the situation below the scale $M_R$, where the PS group is broken to the SM one and the right-handed messengers decouple. The (non-singlet) fields are: the SM fields; the left-messengers $\bar LL+\bar QQ$; 2 right-handed lepton-like components (and their conjugates) in $x_c$ and $F'_c$,$\bar{F_c^\prime}$; 2 right-handed down quark-like components (and their conjugates) in $H_6$ and $F'_c$,$\bar{F_c^\prime}$; linear combinations of the up quark-like and singlet lepton-like components in $\phi$ and $\psi'$ (and their conjugates); linear combination of the quark-doublet like components in $\phi$ and $\Omega$ (and their conjugates); the fields contained in the 5 and $\bar 5$ SU(5) components of $\bar\psi'$, $\psi'$ and $H$. What matters for our purposes is that this also turns out to be a magic field content, with equal contributions to all beta-function coefficients, and thus it is not changing the unification scale.

Below $M_L$, only the SM fields (and possibly a set of full SU(5) multiplets) are supposed to survive. This needs $\ord{M_L}$ $U(1)_R$-breaking mass terms for the linear combinations involving $\psi'$, $\phi$ and $\Omega$ mentioned above, which constitutes a full $10$ and $\overline{10}$ of SU(5). It is not difficult to arrange a superpotential involving a $R=2$ singlet getting a vev at the scale $\ord{M_L} = \ord{\lambda^5\Lambda}$. 

To summarize, we have the following scales: $\Lambda \approx M_\text{GUT} \approx 10^{17}\GeV$, $M_c = 1/R \approx 2\cdot 10^{16}\GeV$, $M_R \approx \lambda^3 \Lambda \approx 2\cdot 10^{15}\GeV$, $M_L \approx \lambda^5\Lambda \approx 3\cdot 10^{14}\GeV$ and a magic field set from $\Lambda$ down to the electroweak scale except for a small threshold. 

\subsection{Neutrinos}
\label{sec:neutrinos}

The light neutrino mass matrix originates from the NR operator $h_{ij} (l'_i h_u)(l'_j h_u)/(2\Lambda_L)$, where $l'_{1,2,3}$ are the three light lepton doublet mass eigenstates: $m^\nu_{ij} = h_{ij} v^2_u/\Lambda_L$. The coefficients $h_{ij}/\Lambda_L$ are obtained by integrating out the $R_P$-odd heavy singlet neutrinos. 

We aim at obtaining a large atmospheric angle $\theta_{23}$, the atmospheric squared mass difference $\dm{23}$ at the correct scale, and the suppression of the solar squared mass difference $\dm{12}$ (in the context of normal hierarchical neutrinos) and of the $\theta_{13}$ angle. Previously \cite{Ferretti:2006df}, the large atmospheric angle and the $\dm{12}/\dm{23}$ suppression were obtained essentially through the single right-handed neutrino dominance mechanism. In fact, the whole idea underlying this flavour model, based on the exchange of a single family of flavour messengers, can be considered as an extension of that mechanism. In order to reproduce the single right-handed neutrino dominance mechanism, the left-handed messengers should have a mass term at the $M_L$ scale (along the $B-L$ direction). Here, we prefer to consider the more economical option in which a such term does not arise or arises at a lower scale. This is interesting also because the large atmospheric mixing arises through a different, unusual mechanism, as we are now going to see. 

In our model, the singlet neutrinos taking part to the see-saw are more numerous than the usual 3. There are in fact 9 $R_P$-odd singlet neutrino fields in the model. These are the usual three right-handed neutrinos $n^c_i$, the SU(2)$_R$ partners of the SM right-handed charged fermions $e^c_i$. In addition, there are $N^c$, $\bar N^c$~\footnote{It is sufficient to consider only the KK zero-modes of $N^c$ and $\bar{N}^c$, because their mass terms arise purely from the PS brane. That means that higher KK mode pairs $(++,--)_{n>0}$ and $(+-,-+)_{n \geq 0}$ decouple from the other fields, because one member of these pairs vanishes at the PS brane and has therefore only a heavy mass term with its partner.}, $A_\Sigma$, and three gauge singlets $S_i$ (additional singlets do not play a role as they have different $R_P$, do not mix with the previous ones, and are not relevant for light neutrino masses). The heavy singlet neutrino mass terms are given by $-(N^c, \bar N^c, A_\Sigma, n^c_i, S_k)^T M_s (N^c, \bar N^c, A_\Sigma, n^c_j, S_h)/2$, where 
\begin{equation}
\label{eq:Msinglets}
M_s = 
\begin{pmatrix}
0 & M_R & \sqrt{\frac{3}{8}}\bar\sigma_c \bar V_c & 0 & b_h M_{SN} \\
M_R & 0 & \sqrt{\frac{3}{8}}\sigma_c V_c & \alpha_j^c v & c_h M_{SN} \\
\sqrt{\frac{3}{8}}\bar\sigma_c \bar V_c & \sqrt{\frac{3}{8}}\sigma_c V^c & M_\Sigma & 0 & 0 \\
0 & \alpha_i^c v & 0 & 0 & a_{ih} M_{Sn} \\
b_k M_{SN} & c_k M_{SN} & 0 & a_{kj} M_{Sn} & 0 
\end{pmatrix}
\end{equation}
and the light neutrino mass operator is
\begin{multline}
\label{eq:seesaw0}
\frac{h_{ij}}{2\Lambda_L} (l'_i h_u)(l'_j h_u) = \frac{1}{2}\left[ (M^{-1}_s)_{N^cN^c} (\lambda^c_2 l'_2)^2
+(M^{-1}_s)_{n^c_3n^c_3} (\lambda_3l'_3 )^2
+2(M^{-1}_s)_{N^cn^c_3} (\lambda^c_2 l'_2) (\lambda_3 l'_3) \right] h^2_u ,
\end{multline}
so that
\begin{equation}
\label{eq:lightnu}
m_\nu = v^2_u
\begin{pmatrix}
0 & 0 & 0 \\
0 & (\lambda^c_2)^2 (M^{-1}_s)_{N^cN^c} & \lambda^c_2\lambda_3 (M^{-1}_s)_{N^cn^c_3} \\
0 & \lambda^c_2\lambda_3 (M^{-1}_s)_{N^cn^c_3} & \lambda_3^2 (M^{-1}_s)_{n^c_3n^c_3}
\end{pmatrix} .
\end{equation}
The entries in the first row and column, accounting for the solar and $\theta_{13}$ mixing angles, will be generated, as in the case of charged fermion masses, by higher order operators, possibly controlled by a flavour symmetry. In \eq{Msinglets} the entries set to zero arise at a negligible level.

In order to get a large atmospheric mixing angle from \eq{lightnu}, we need $(M^{-1}_s)_{N^cN^c} \sim (M^{-1}_s)_{N^cn^c} \sim (M^{-1}_s)_{n^cn^c}$ and in order to obtain the (mild) hierarchy between the solar and atmospheric squared mass differences, we need the determinant $(M^{-1}_s)_{N^cN^c} (M^{-1}_s)_{n^cn^c} -  (M^{-1}_s)_{N^cn^c}^2 $ to be suppressed. This can be obtained if $M_{SN}\sim M_{Sn} > M_L$, in which case
\globallabel{eq:inverseM}
\begin{gather}
(M^{-1}_s)_{N^cN^c} \sim (M^{-1}_s)_{N^cn^c_3} \sim (M^{-1}_s)_{n^c_3n^c_3} \sim  \frac{1}{2 M_R}  \mytag, \\
(M^{-1}_s)_{N^cN^c} (M^{-1}_s)_{n^c_3n^c_3} -(M^{-1}_s)_{N^cn^c_3}^2 \sim  \frac{M^2_R}{V_c^2} (M^{-1}_s)_{N^cN^c}^2 . \mytag
\end{gather}
This is indeed what our model gives: $M_{SN}\sim M_{Sn} \sim \lambda M_R > \lambda^2 M_R \sim M_L$ and $M_R/V_c \sim\lambda < 1$. 

Taking into account all ${\cal O}(1)$ coefficients we finally obtain for the light neutrino masses and the atmospheric mixing
\globallabel{eq:neutrinopredictions}
\begin{gather}
m_3 =   \frac{v_h^2}{M_R} \frac{A}{2 \sin^2 \theta_{23}} \mytag, \\
\frac{m_2}{m_3} =  \frac{4 \lambda^2}{3} \sin^2 2\theta_{23} B \mytag, \\[1mm]
\tan \theta_{23} =  C, \mytag
\end{gather}
where
\globallabel{eq:neutrinoO1}
\begin{gather} 
A =  \frac{ (\lambda_2^c)^2 \sigma_c}{\bar{\sigma}_c}, \qquad
B =  \frac{\bar{\sigma}_c x^2}{\sigma_c y^2}, \qquad
C =  \frac{ \lambda_2^c \sigma_c \det a}{\lambda_3 y} \mytag,\\[2mm]
x =  c_2 \left( a_{12} a_{31} - a_{11} a_{32} \right) + c_3 \left( a_{11} a_{22} - a_{12} a_{21} \right) \mytag, \\[4mm]
y =  \bar{\sigma}_c x - \sigma_c b_3  \left( a_{11} a_{22} - a_{12} a_{21} \right). \mytag
\end{gather}
In order to agree with the experimental values  $m_2/m_3  \approx  \sqrt{\dm{12}/\dm{23}}   \sim  0.2 $ and $\tan \theta_{23} \sim 1$, one has to require that the above functions of ${\cal O}(1)$ coefficients take the (reasonable) values $B  \sim  3$ and $C  \sim 1$. The atmospheric squared mass difference provides an experimental determination of $m_3 \approx \sqrt{\dm{23}}$, which translates into a determination of the scale $M_R$, given by
\begin{equation}
M_R = \frac{v_h^2 A}{2 \cos^2 \theta_{23} \sqrt{\dm{23}}} \sim A \times 6 \times 10^{14} {\rm GeV}. 
\end{equation}
To achieve agreement with the numerical determination of the various scales provided by gauge  coupling unification, we have to require that $A \approx$ 3. 

\subsection{Gauge coupling unification}
\label{sec:unification}

In the context of the 5D model under consideration, gauge coupling unification has novel features. A generic feature of 5D models is the presence of threshold effects associated to the tower of KK excitations associated to bulk fields (sometimes useful to improve the gauge coupling unification prediction). In our case such thresholds do not arise at the 1-loop level. 

The reason why the KK threshold corrections are typically present in 5D models is that the very breaking of the unified group by boundary conditions also splits the multiplets associated to the fields in the bulk into submultiplets that are not full representations of the gauge group and have different masses. In our case, however, such non-full representations happen again to preserve MSSM 1-loop unification at each floor of the KK tower (and do not affect the unification scale). 

The beta function coefficients at one loop, neglecting the threshold effects associated to $V_c > M_R$, follow. Below $M_L$ the coefficients are the MSSM ones: $(b_1,b_2,b_3) = (33/5,1,-3)$. The quantity determining the $\alpha_s$ prediction at one loop is $(b_3-b_2)/(b_1-b_2) = -5/7$. In our case, the content of fields between $M_L$ and $M_R$ gives a common shift to the coefficients,
even if the new fields are not in full SU(5) multiplets, so that the condition above is trivially satisfied. The coefficients are $(b_1,b_2,b_3) = (78/5, 10, 6)$. The 1-loop MSSM $\alpha_3$ prediction is preserved independently of the value of the scale $M_L$. Above $M_R$ the gauge group is $G_\text{PS}$, the SM gauge couplings are matched into the PS ones by
\begin{equation}
\label{eq:gaugematching}
\frac{1}{\alpha_4} = \frac{1}{\alpha_3},
\qquad
\frac{1}{\alpha_L} = \frac{1}{\alpha_2},
\qquad
\frac{1}{\alpha_R} = \frac{5}{3} \frac{1}{\alpha_1} -\frac{2}{3} \frac{1}{\alpha_3},
\end{equation}
and the 1-loop MSSM $\alpha_3$ prediction is preserved if $(b_4-b_L)/(b_R-b_L) = -1/3$. This is indeed what happens. In fact, between $M_R$ and $1/R$ we have $(b_L,b_R,b_4) = (28, 34, 26)$ and above $1/R$ each KK set adds $(b_L,b_R,b_4) = (10,10,10)$. The coefficients associated to the KK floors are the same for the three gauge couplings but still they do not correspond to full SU(5) multiplets. The 4D gauge couplings are supposed to unify at the cutoff $\Lambda$ within an intrinsic uncertainty due to unknown brane Yang-Mills terms\cite{Hall:2001pg,SU5OFGUTS}. The latter are suppressed because of the strong coupling regime assumption. The correction to $g^2(\Lambda)$ is of order $1/l^V_4$ and is therefore expected to be a few percent. 

\begin{figure}[h]
\begin{center}
\includegraphics[angle=-90,width=0.49\textwidth]{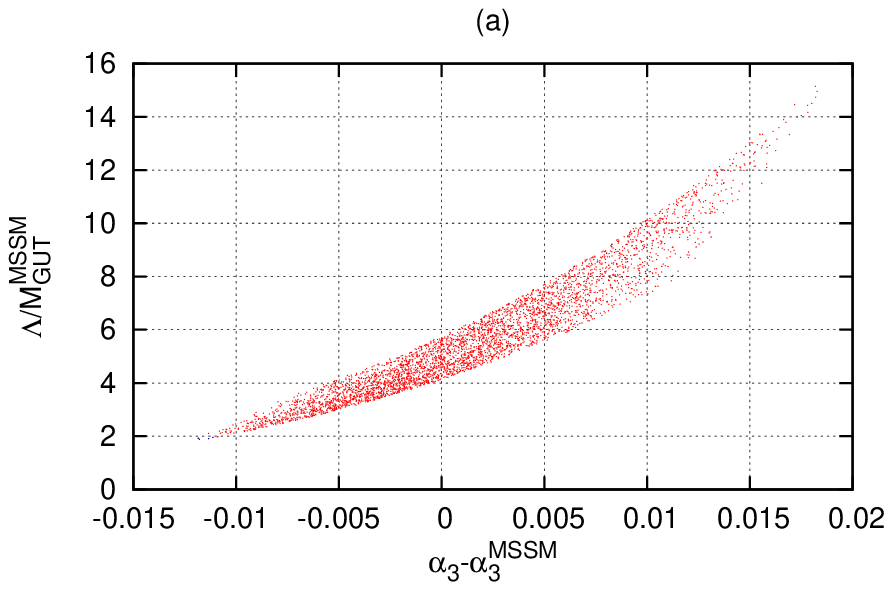}
\includegraphics[angle=-90,width=0.49\textwidth]{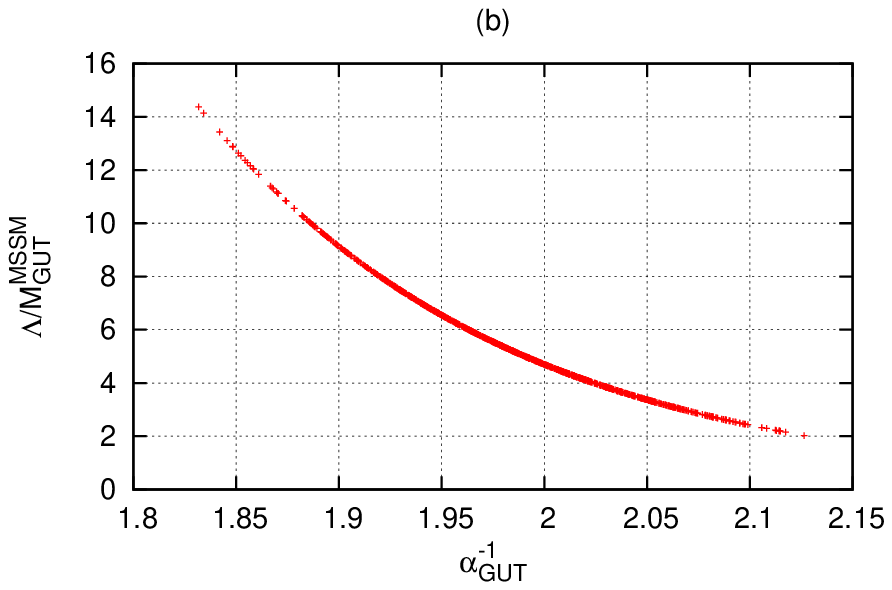}
\end{center}
\caption{Correlation between the threshold effects on $\alpha_s(M_Z)$ and $\Lambda$ (left). Value of $\alpha_\text{GUT}$ at the cutoff scale (right).}
\label{fig:thresholds}
\end{figure}

We are now in the position to calculate the prediction for $\alpha_s(M_Z)$ and $\Lambda = M_\text{GUT}$ in our model and to discuss perturbativity, which we do in the limit in which the brane corrections vanish. We quote our results in terms of $\alpha^{-1}_s(M_Z) - \alpha^{-1}_s(M_Z)^\text{MSSM}$ and $\Lambda/M_\text{GUT}^\text{MSSM}$, as the latter quantities can be calculated with a good accuracy at the one loop level (the MSSM thresholds and the 2-loop effects below $M_L$ mostly cancel). Neglecting high energy thresholds due to small differences among the masses of the fields that were assumed to be at the $M_L$, $M_R$, or $n/R$ scale, we have $\alpha_s(M_Z) = \alpha_s(M_Z)^\text{MSSM}$ and $\Lambda = (M_\text{GUT}^\text{MSSM})^2 /M_R$. The largest threshold effects come from the fact that some of the fields actually live at the scale $V_c > M_R$ or slightly below. More precisely, there are two additional thresholds corresponding to the non singlet PS/SM vector bosons. Their masses are denoted by $M_E$ (for the SU(2)$_R$ extra bosons) and $M_U$ (for the SU(4)$_c$ extra bosons). Numerically such splittings are approximately $M_{E}/M_R \sim 2$, $M_{U}/M_R \sim 1.2$. 
In the presence of such thresholds we have: 
\globallabel{eq:thresholds}
\begin{gather}
\alpha^{-1}_s(M_Z) - \alpha^{-1}_s(M_Z)^\text{MSSM} = 
\frac{45}{14\pi} \log \frac{M_U}{M_R} +
\frac{18}{14\pi} \log \frac{M_E}{M_R} -
\frac{30}{14\pi} \log \frac{V_c}{M_R} \mytag, \\
\Lambda = \frac{(M^\text{MSSM}_\text{GUT})^2}{M_R} \left(\frac{V_c}{M_R}\right)^{8/7} \left(\frac{M_R}{M_U}\right)^{12/7} \left(\frac{M_R}{M_E}\right)^{9/7} . \mytag
\end{gather}
The correlation between the threshold effects in $\alpha_s$ and $\Lambda$ is shown Figure~\ref{fig:thresholds}a, where the predictions for $\Lambda/M_\text{GUT}^\text{MSSM}$
and $\alpha_3(M_Z) - \alpha^{\rm MSSM}_3 (MZ)$ are plotted for $M_E$ and $M_U$ randomly chosen in the ranges $1 \leq M_E/M_R \leq 3$, $1 \leq M_U/M_R \leq 3$ and $V_c/M_R$ is varied by 50\% around the central value $1/\lambda$ (all with uniform distribution in logarithmic scale). We used $\Lambda/M_R = 0.5/\lambda^3$, $M_R/M_L = 0.5/\lambda^2$ and $\Lambda R =4$. The figure shows that it is possible to reduce the MSSM prediction for $\alpha_s(M_Z)$ while keeping $\Lambda$ above $M_\text{GUT}^\text{MSSM}$, which helps with proton decay, as we will see in Section~\ref{sec:protondecay}. Note that a few KK masses $n/R$ enter the running before the cutoff $\Lambda$ is reached. Given the presence of so many new degrees of freedom, it is important to check that the theory (in particular the running of gauge couplings) remains calculable up to the cutoff scale. This is indeed the case, as shown in Figure~\ref{fig:thresholds}b. 

\subsection{Proton decay}
\label{sec:protondecay}

In 4D GUTs the leading contributions to proton decay usually come from d=5 operators (via Higgs triplet exchange), while d=6 operators (via extra gauge boson exchange) are subleading. In 5D orbifold GUT models one generally finds the opposite picture: d=5 operators are strongly suppressed due to the U(1)$_R$ symmetry, while d=6 operators are more important because the mass of the extra gauge bosons is smaller than in 4D GUTs $(M_c < M^\text{MSSM}_\text{GUT})$. This general picture also holds in our model. The dominant contribution to proton decay is SO(10)/PS gauge boson exchange, leading to a lower bound on the compactification scale $M_c > \ord{10^{16}} \GeV$. Before deriving this bound, we briefly comment on the suppression of the d=5 operators.

The low energy effective theory below $M_L$ is the MSSM with R-parity. This allows only for two d=5 operators that induce proton decay (originating from exchange of chiral superfields with standard model quantum numbers $(3,1)_{-1/3}$)\cite{Sakai:1981pk},
\begin{equation}
\label{eq:poperators}
QQQL |_{\theta^2} \qquad \text{and} \qquad
D^c U^c U^c E^c |_{\theta^2}.
\end{equation}
The operators in \eq{poperators} are suppressed because they violate the U(1)$_R$ symmetry. Despite there is a source of U(1)$_R$ breaking at the scale $M_L$ (which is needed to provide the mass term $\psi^\prime \bar{\psi^\prime}$), it violates U(1)$_R$ by +2 units and hence cannot induce the above operators. Therefore they can be assumed to be generated at the SUSY breaking scale and hence to be strongly suppressed. For the case of a minimal SUSY breaking sector (i.e. just involving a spurion with F-term vev and $R=0$), the d=5 operators are clearly subleading. 

Let us now consider the d=6 operators inducing proton decay. There are two such operators in the MSSM: 
\begin{equation}
\label{eq:poperators2}
Q Q (U^c)^\dagger (E^c)^\dagger    |_{\theta^4} = u_L d_L \bar{u}_R^c \bar{e}_R^c 
\qquad\text{and}\qquad 
Q L (U^c)^\dagger (D^c)^\dagger    |_{\theta^4} = u_L e_L \bar{u}_R^c \bar{d}_R^c .
\end{equation}
These operators arise from the exchange of gauge fields with quantum numbers $(3,2)_{(5/6)}$ and  $(3,2)_{(1/6)}$, which are contained in the decomposition of the SO(10)/PS gauge bosons $V_{+-}^{(226)}$. Therefore the suppression of the d=6 operators is $\ord{1/M_c^2}$. The numerical coefficients can be calculated in close analogy to the SU(5) case \cite{Hebecker:2002rc}, as we are going to show in the following. 

We start with the relevant part of the 5d action
\begin{equation}
S_5 =  \int d^4 x \int_0^{\pi R/2} dy \left( \frac{1}{2 g_5^2} {\rm Tr} \left( W^2 |_{\theta^2} + {\rm h.c.} + 4 \partial_5 V \partial_5 V |_{\theta^4} \right) + \delta(y) \psi_i^\dagger e^{2V} \psi_i |_{\theta^4} \right),
\end{equation}
which fixes the normalization in the KK expansion of $V_{+-}^{(226)}$:
\begin{equation}
 V_{+-}(x,y) = \sqrt2 \sum_{n=0}^{\infty} V_{+-}^{(2n+1)}(x) \cos{\frac{(2n+1)y}{R}}.
\end{equation}
The 4D effective Lagrangian is given by
\begin{equation}
{\cal L}_4 = \frac{2}{g_4^2} \sum_{n=0}^{\infty} \left(\frac{2n+1}{R}\right)^2 {\rm Tr} \left( V_{+-}^{(2n+1)} V_{+-}^{(2n+1)} \right) + 2 \sqrt 2 \left( \psi_i^\dagger \sum_{n=0}^{\infty}  V_{+-}^{(2n+1)} \psi_i \right) |_{\theta^4}
\end{equation}
where we have used $1/g_4^2 = (\pi R/2)/g_5^2$. Integrating out the heavy gauge bosons, we obtain the following effective Lagrangian containing the operators in \eq{poperators2} 
\begin{equation}
{\cal L}_4 = \frac{g_4^2}{M_c^2} \frac{\pi^2}{4} \left( 2 u_l d_l \bar{u}^c_R \bar{e}^c_R + 2 u_l e_l \bar{u}^c_R \bar{d}^c_R \right) ,
\end{equation}
which is the result for an ordinary 4D SO(10) GUT with gauge boson mass given by $M_c$, enhanced by the $\pi^2/4$ factor, which arises from summing over KK states. The proton decay rate into $\pi^0 e^+$ follows \cite{Hisano:1992jjHisano:2000dg}:
\begin{equation}
\Gamma(p \rightarrow \pi^0 e^+) = 8 \alpha_H^2 \frac{\pi^4}{16} \left( \frac{g_4(M_c)^2 A_R}{M_c^2} \right)^2 \frac{m_p}{64 \pi f_\pi^2} \left( 1 + D + F \right)^2 . 
\end{equation}
Using the hadronic parameter $\alpha_H=0.015$ ${\rm GeV}^3$, the pion decay constant $f_\pi = 0.13$ ${\rm GeV}$ and the chiral perturbation theory parameters $D= 0.80$ and $F=0.47$, the partial lifetime can be estimated to be
\begin{equation}
1/\Gamma(p \rightarrow \pi^0 e^+) \approx 2.0 \times 10^{33}  \left( \frac{\alpha_4 (M_c)}{1/12} \right)^{-2} \left( \frac{A_R}{2.5} \right)^{-2} \left( \frac{M_c}{10^{16} {\rm GeV}} \right)^4 \; \text{years}, 
\end{equation}
where the normalization uses the typical values of the gauge coupling at $M_c$ and of the renormalization coefficient $A_R$ we obtain in the model. A comparison with the PDG bound on the partial lifetime \cite{Shiozawa:1998si} 
\begin{equation}
1/\Gamma(p \rightarrow \pi^0 e^+) > 1.6 \times 10^{33} {\rm years}, 
\end{equation}
gives the limit we anticipated for the compactification scale 
\begin{equation}
M_c >  9.5 \times 10^{15}\GeV \left( \frac{\alpha_4 (M_c)}{1/12} \right)^{1/2} \left( \frac{A_R}{2.5} \right)^{1/2}. 
\end{equation}
For $M_c \sim 2 \times 10^{16}$ ${\GeV}$, 
the lifetime turns out to be
\begin{equation}
1/\Gamma(p \rightarrow \pi^0 e^+) \sim 3 \times 10^{34} \text{years},
\end{equation}
within the sensitivity of future experiments designed to reach a limit on the partial lifetime of $\ord{10^{35}}$ years \cite{Shiozawa:1999yg}.

\subsection{The first family}
\label{sec:first}

We have seen that most features of the SM fermion spectrum concerning the second and third fermion families can be accounted for without flavour symmetries or other dynamics related to the fermion family indices. The smallness of the first family charged fermion masses with respect to two heavier ones is also explained. On the other hand, in order to account for the quantitative aspects of the spectrum associated to the first family, it may be necessary to introduce some type of flavour dynamics. In this Section, we sketch the case of a flavour symmetry that only acts on the first family, the second and third families being not charged. 

Before showing one example of such a symmetry, let us discuss what type of Yukawa texture it should provide. One important guideline on the first family entries of the Yukawa matrices is given by the constraints from FCNC and LFV effects. Such effects are induced by off-diagonal elements in the sfermion mass matrices written in the so-called SCKM basis, the basis in which the Yukawa matrices of the corresponding fermions are diagonal. For example, \eq{soft} gives 
\begin{equation}
\begin{gathered}
(m^2_{RR})_\text{SCKM} =  U_R \begin{pmatrix}
(m^2_{\tilde{e^c}})_{11} & 0 & 0 \\
0 & (m^2_{\tilde{e^c}})_{22}& (m^2_{\tilde{e^c}})_{23} \\
0 & (m^2_{\tilde{e^c}})_{23} & (m^2_{\tilde{e^c}})_{33}
\end{pmatrix} U^\dagger_R, \\
(m^2_{LL})_\text{SCKM} =  U_L \begin{pmatrix}
(m^2_{\tilde{l}})_{11} & 0 & 0 \\
0 &  (m^2_{\tilde{l}})_{22} & 0 \\
0 & 0 & m^2_{\tilde{L}}
\end{pmatrix} U^\dagger_L 
\end{gathered}
\end{equation}
in the slepton sector, where $U_L$ and $U_R$ are the unitary matrices diagonalizing the charged lepton Yukawa matrices, $Y^E = U^\dagger_R Y^E_\text{diag} U_L$. Because of the sizable non-degeneracy between $m^2_{11}$ and $m^2_{22}$, see \eq{llog}, a small rotation between the first two families in $U_L$ or $U_R$ induces a sizable off-diagonal 12 entry in $m^2_\text{SCKM}$. This in turn gives sizable contributions to the $\mu\to e\gamma$ branching ratio. Therefore, the matrices $U_L$ and $U_R$ should involve small rotations in the ``12'' block, which translates into small 12 and 21 entries in the Yukawa matrices. Analogous considerations hold in the down quark sector because of the constraints from the Kaon system ($\Delta M_K$ and $\epsilon_K$). 

Because of the above constraints and of the stronger hierarchy $m_u/m_t \ll m_d/m_b$, i) the $d$-quark and electron masses must originate from $Y^D_{11}$ and $Y^E_{11}$ respectively, ii) the corresponding up quark Yukawa should be smaller, $Y^U_{11} \ll Y^{D,E}_{11}$, and iii) the $V_{us}$ angle must originate from the $Y^U_{21}$, with $Y^D_{12,21}$ and $Y^E_{12,21}$ smaller than the corresponding 11 elements. 

Let us now see how the above requirements can be satisfied. The Yukawa entries $Y^{D,E}_{11}$ can be generated by the operator $\hat \psi_1\hat \psi_1 \hat \phi \hat h \hat Z$ in the SO(10)-brane superpotential, where $Z$ is an SO(10) adjoint living on the SO(10) brane and taking a vev in the SU(5)-invariant direction, with $Z_{24}$ charge 12, $R$-charge zero, and $\big\langle{\hat Z}\big\rangle \sim \lambda^2$. After plugging the vevs of $\phi$ and $Z$ and using \eqs{couplings} to write the operator in terms of canonically normalized fields, we get $Y^{D,E}_{11} \sim \lambda^5$, which is about what needed\footnote{A term involving 2 brane fields and 1 bulk field is enhanced with respect to a term involving 1 brane field and 2 bulk fields (which is assumed to be $\ord{1}$) by a factor $1/\lambda$, when expressed in terms of canonically normalized fields.\label{foo:enhancement}}. The interesting property of the operator $\hat \psi_1\hat \psi_1 \hat \phi \hat h \hat Z$ is that it does not give rise to a contribution to $Y^U_{11}$, as desired. The reason can be understood as follows. The operator $\hat \psi_1\hat \psi_1 \hat \phi \hat h$ does not contribute to any SM Yukawa because the vev of $\phi$ in the $B-L$ direction makes it antisymmetric in the two $\psi$'s. In the case of the up quark Yukawas, the antisymmetry is not spoiled by the presence of $\hat Z$. This is because both the right-handed and left-handed up quarks belong to the same representation of SU(5), so that the vev of $Z$ factorizes. Therefore the up quark Yukawa still vanishes when $\hat Z$ is included in the operator. On the other hand, in the case of the down quark and charged lepton Yukawas, the antisymmetry of $\psi_1\hat \psi_1 \hat \phi \hat h$ is spoiled by the presence of $\hat Z$, as the right-handed and left-handed fields in this case belong to different SU(5) representations, and the vev of $Z$ cannot be factorized. Therefore, the corresponding Yukawas do not vanish when $\hat Z$ is included in the operator. As for $V_{us}$, we said above that it should be generated by the $Y^U_{21}$ matrix element. The latter can be generated at the correct level by the operator $\hat\Sigma\hat\psi_1 \hat h \hat F'_c \hat Z$. This operator involves fields belonging to two different  branes and can be generated by the exchange of bulk fields with no zero modes (specifically the SO(10) partner of $F$ and its SU(2)$_R$ conjugated partner) along the lines of footnote~\ref{foo:branebrane}. Once the vevs of $\hat F'_c$ and $\hat Z$ have been plugged, $\Sigma$ has been substituted by its light ${u^c_2}'$ component (from eq.~(\ref{eq:masses}c) one finds that the latter is suppressed by a factor $\lambda^3$, $\bar T_\Sigma \sim \lambda^3 {u^c_2}' + \text{heavy}$), and the ``brane-brane-bulk'' enhancement in footnote~\ref{foo:enhancement} has been taken into account, one gets $\lambda^U_{21} \sim \lambda^5$, which is about what needed to account for $V_{us}$. 

As discussed above, if the operator $\hat \psi_1\hat \psi_1 \hat \phi \hat h \hat Z$ was accompanied by the operator $\hat \psi_2\hat \psi_1 \hat \phi \hat h \hat Z$, potentially dangerous FCNC and LFV effects could be generated. In order to be on the safe side and forbid the second operator one can consider a U(1) flavour symmetry under which $\psi_{2,3}$ are not charged, as promised, and $\psi_1$ has charge 1. Giving the $Z$ field U(1) charge -2 allows the first operator while forbidding the second one. Such a symmetry would also forbid the $\hat\Sigma\hat\psi_1 \hat h \hat F'_c \hat Z$ operator, as there is no U(1) assignment to $\Sigma$ compatible with \eq{Wflavour}. In order to use the U(1) symmetry above one should then use a different operator, $\hat\Sigma\hat\psi_1 \hat h \hat F''_c \hat Z$, involving a new field $F''_c$ on the PS brane having same gauge, $Z_{24}$, U(1)$_R$ charges but U(1) charge 1 (accompanied by a corresponding $F$ field in order not to spoil unification and by the conjugated fields). 

Finally, the operator $\hat \psi_1\hat \psi_1 \hat \phi \hat h \hat Z$ gives the wrong $m_e/m_d$ ratio. The latter can be fixed if the first family left-handed or right-handed leptons mix. A mixing involving first family left-handed leptons can also account for the solar angle and the other features of the neutrino mass matrix. 

\subsection{Summary of Section~\ref{sec:SO(10)}}
\label{sec:summ}

In this section, we embedded the PS model of Section~\ref{sec:lowE} into a supersymmetric 5D orbifold SO(10) model. In particular, the two discrete symmetries $\mathbf{Z}_2$ and $R_P$ are embedded into a larger discrete group $Z_{24}$ and into a continuous $R$-symmetry $U(1)_R$ respectively. New fields are introduced in order to generate the proper vevs, to account for the neutrino sector and to preserve unification. We proceeded with an analysis of this setup as follows
\begin{itemize}
\item We studied the possibility that the $\ord{1}$ parameters and some of the mass scale hierarchies originate from a strong coupling regime through an order parameter $\lambda \approx 0.24$, in agreement with naive dimensional analysis. 

\item We wrote down the most general superpotential consistent with the symmetries, where we included two anomalously small coefficients, which represent, together with the strong coupling hierarchies, the seeds of all hierarchies of the model. In particular, the low-energy fermion mass hierarchies and small mixings have their very origin in this assumption. 

\item 
The mass scales of the model in Section~\ref{sec:lowE}, $M_R$ and $M_L$, and the compactification scale $M_c$ turn out to be related to the cutoff scale $\Lambda \approx 10^{17}\GeV$ by powers of the order parameter $\lambda$ up to $\ord{1}$ coefficients. The field content at every scale is such that the MSSM unification of gauge couplings is preserved (``magic field sets''), but takes place at the higher scale $\Lambda$. 

\item In the neutrino sector we find good agreement with the data up to $\ord{1}$ coefficients without making further assumptions. In particular, we obtain a large atmospheric mixing by means of an unusual mechanism. 

\item We studied gauge coupling unification in more detail to obtain a prediction for $\alpha_s(M_Z)$ and the (new) GUT scale $\Lambda$ as a function of the small thresholds of the model. Depending on the $\ord{1}$ uncertainties of such thresholds, we find that  it is possible to obtain a prediction for $\alpha_s(M_Z)$ that improves on the standard MSSM prediction. 

\item We computed the proton lifetime resulting from our unified setup. The contribution to the decay rate from d=5 operators is strongly suppressed, which is a well-known consequence of the U(1)$_R$ symmetry. The contribution from d=6 operators leads to a proton partial lifetime $1/\Gamma(p \rightarrow \pi^0 e^+) \sim 3 \times 10^{34}$ years, which is consistent with the present bound from SuperKamiokande  and within the reach of future experiments.

\item The further suppression of the first family charged fermion masses is also a consequence of the model. On the other hand, in order to account for the quantitative aspects of the spectrum associated to the first family, it may be necessary to introduce some type of flavour dynamics. We briefly sketched a possible implementation of this possibility. 
\end{itemize}

\section{Conclusions}

Following~\cite{Ferretti:2006df}, we illustrated a neutrino-inspired supersymmetric model of fermion masses and mixings. As in the case of the second and third neutrinos in the context of the single right-handed dominance mechanism, the dominant exchange of a single set of left-handed messengers, with unconstrained, $\ord{1}$ couplings accounts for most features of the masses and mixings of the second and third families and for the main qualitative features associated to the first family. While some specific flavour dynamics may need to be invoked for a quantitative description of the first family, it is impressive that so many other features do not actually need a flavour symmetry or other dynamics to be explained. Among them, we cite the suppression of the mass of the second family of charged fermions with respect to the third one, $|V_{cb}| \sim m_s/m_b$, $(m_\tau/m_b)_\text{GUT} \approx 1$, $(m_\mu/m_s)_\text{GUT} \approx 3$, the larger suppression in the up quark sector, $m_c/m_t\ll m_s/m_b$, the further suppression of the first charged fermion family mass. 

In particular, we discussed the flavour phenomenology and the possibility to embed the model in a Grand Unified setting, preserving gauge coupling unification. 

The interactions of the MSSM fields with the left-handed and right-handed flavour-messengers living at energies much higher than the electroweak scale leave their imprint on the sfermion soft masses through radiative effects, which induces FCNC and LFV effects at low energy. The peculiar feature of our model is that the Yukawa interactions of the flavour messengers with all the three light families are described by $\ord{1}$ couplings, leading to sizable flavour-violating effects in the sfermion mass matrices. As the model targets especially the second and third families, we concentrated on the $\tau\to\mu\gamma$ decay. The present limit $\text{BR}(\tau\to\mu\,\gamma)\lesssim 6.8\cdot 10^{-8}$ only gives a weak constraint on the supersymmetry parameter space, while a Super $B$-Factory able to reach a sensitivity of $10^{-9}$ would test a large portion of it. The structure of the radiatively-induced contributions to the soft masses in the moderate $\tan\beta$ regime,
$\tilde m^2_{ij} = m^2_0 \delta_{ij} + \sigma c_{ij} m^2_0$, is particularly promising for detecting LFV at colliders. This structure gives rise to large mixing between the second and third family independent of how small $\sigma$ is. By making $\sigma$ small the branching ratio $\text{BR}(\tau\to\mu\,\gamma)$ can be kept under control, but the LFV effects at colliders do not get suppressed, as long as $|\tilde m_2 - \tilde m_3|$ does not get too small, because the mixing angle remains large in the $\sigma\to 0$ limit. 

The model can be embedded in a SO(10) supersymmetric GUT in 5 dimensions. As usual in 5D GUTs, the doublet-triplet splitting and the suppression of dimension five operators contribution to proton decay can be easily obtained in terms of orbifold boundary conditions and an $R$-symmetry. The three MSSM families are embedded into the spinorial representation of SO(10), so that the successful SO(10) predictions of the SM fermion gauge quantum numbers is maintained. We exhibited a full model accounting for the vevs used in the generation of the flavour structure, among the other things. 

Two nice features of the model we are proposing are the possibility to relate the $\ord{1}$ parameters to strong couplings and the fact that one of the small order parameters arises from the strong coupling assumption, in particular from the loop factor $l_4$ in 4 dimensions, $\lambda \equiv 1/l_4^{1/4} \approx 0.24$. 

Gauge coupling unification is obtained in 5D in a novel way. At each scale up to the cutoff $\Lambda \sim 10^{17}\GeV$  the field content has the ``magic'' property that it exactly preserves the MSSM 1-loop successful prediction for $\alpha_s$, despite the field content does not constitute a full SU(5) representation~\cite{CFRZ}. On the other hand the unification scale, here identified with the cutoff $\Lambda$, gets larger, thus helping to avoid a too fast proton decay rate. This is also true in the energy range between the compactification scale and the cutoff, where the KK towers of states enter the running of gauge couplings. This is quite unusual. A generic feature of 5D GUT models is in fact the presence of threshold effects associated to the tower of KK excitations associated with bulk fields (sometimes useful to improve the gauge coupling unification prediction). The latter are associated to the very mechanism of GUT breaking by orbifold boundary conditions, which splits the GUT multiplets in the KK tower. In our case, such thresholds do not arise at the 1-loop level, because the GUT multiplets are split into two ``magic'' sets of fields preserving the MSSM gauge coupling unification. 

As for proton decay, dimension-five operators are strongly suppressed due to a U(1)$_R$ symmetry, while d=6 operators are important because the mass of the extra gauge bosons is not too large. The lower limit on the proton lifetime gives a lower limit on the compactification scale $M_c \gtrsim  9.5 \times 10^{15}\GeV$, which is compatible with a larger unification scale of $\Lambda \sim 10^{17}\GeV$. 

Some interesting features also arise in the neutrino sector. Previously~\cite{Ferretti:2006df}, the large atmospheric angle and the $\dm{12}/\dm{23}$ suppression were obtained essentially through the single right-handed neutrino dominance mechanism, which in fact motivates the idea underlying this flavour model. Here, we consider another, more economical option in which the large atmospheric mixing arises through a different, unusual mechanism. In our model, which involves in the see-saw more than the usual 3 singlet neutrinos (9, in fact), the 22, 23, 32, 33 entries of the light neutrino mass matrix turn out to be proportional to four matrix elements of the inverse heavy singlet neutrino mass matrix. Such elements turn out to be of the same size and their determinant turns out to be suppressed because it is related to elements of the singlet neutrino mass matrix. The atmospheric squared mass difference is determined, up to an $\ord{1}$ factor, to the scale at which the flavour messengers live, which in turn is given in terms of the unification scale by \eq{MR}. We then get a large atmospheric angle $\theta_{23}$, the atmospheric squared mass difference $\dm{23}$ at the correct scale, and the suppression of the solar squared mass difference $\dm{12}$ (in the context of normal hierarchical neutrinos) and of the $\theta_{13}$ angle. 

The origin of the peculiar pattern of SM fermion masses and mixings might be related to the breaking of a flavour symmetry, or to localization in extra dimensions, or it could be entangled to the structure of string theory. All in all, we find interesting that so many features of the fermion spectrum could instead just be due to the relative lightness (and structure) of a single family of flavour messengers, with ``horizontal'' flavour dynamics playing essentially no role. 

\section*{Acknowledgments}
We thank Marco Serone for many useful discussions and Pietro Slavich for helpful advice with {\tt SusyBSG}. This work was supported by the European Network of Theoretical Astroparticle Physics ILIAS/N6 (contract RII3-CT-2004-506222). 

\newpage
\section*{Appendix}
In this Appendix we present the 1-loop Renormalization Group Equations (RGE) of the effective model in Section~\ref{sec:lowE} above and below the scale $M_R$.

\subsection*{RGEs above $M_R$}
The relevant superpotential above $M_R$ is given by:
\begin{multline}
W_{\rm PS} =
\lambda_i f^c_i F h + \lambda^c_i f_i F^c h +\alpha_i\phi f_i\bar F+\alpha^c_i\phi f^c_i\bar F^c + a\bar F^c X_c F^c  + \bar\sigma_c \bar F'_c \Phi F^c +\sigma_c\bar F^c\Phi F'_c, \nonumber
\end{multline}
while the SUSY breaking potential reads
\begin{align}
\begin{split}
V_{\rm SSB} &= (m^2_f)_{ij}\tilde{f}^*_i \tilde{f}_j +
 (m^2_{f^c})_{ij}\tilde{f^c_i}^* \tilde{f^c_j} + m^2_h \tilde{h}^* \tilde{h}
+ m^2_\phi \tilde{\phi}^* \tilde{\phi} + m^2_F \tilde{F}^* \tilde{F}
+ m^2_{\bar{F}} \tilde{\bar{F}}^* \tilde{\bar{F}} 
+ m^2_{F^c} \tilde{F^c}^* \tilde{F^c}\nonumber \\
&
+ m^2_{\bar{F^c}} {\tilde{\bar{F^c}}}^* \tilde{\bar{F^c}} 
+ m^2_{F_c^\prime} \tilde{F^\prime_c}^* \tilde{F^\prime_c}
+ m^2_{\bar{F^\prime_c}} {\tilde{\bar{F^\prime_c}}}^* \tilde{\bar{F^\prime_c}} 
+ m^2_\Phi \tilde{\Phi}^* \tilde{\Phi}
+ m^2_X \tilde{X}^* \tilde{X}
+ m^2_{X^c} \tilde{X^c}^* \tilde{X^c}
+ m^2_H \tilde{H}^* \tilde{H}\nonumber \\ 
&
+ \left( A^\lambda_i \tilde {f^c}_i \tilde{F} \tilde{h}
+ A^{\lambda^c}_i \tilde{f}_i \tilde{F^c} \tilde{h}
+ A^\alpha_i \tilde{\phi} \tilde{f}_i \tilde{\bar{F}}
+ A^{\alpha^c}_i \tilde{\phi} \tilde{f^c}_i \tilde{\bar{F^c}} 
+ A^a \tilde{X^c} \tilde{\bar{F^c}} \tilde{F^c} +A^{\sigma_c} \tilde{\bar{F^c}} \tilde{\Phi} \tilde{F}^\prime_c
+ A^{\bar{\sigma}_c} \tilde{\bar{F_c^\prime}} \tilde{\Phi} \tilde{F^c}
+ \text{h.c.} \right)
\nonumber \\
& +\frac{1}{2}\left(M_4 \tilde{W}_4 \tilde{W}_4 + M_L \tilde{W}_L \tilde{W}_L
 + M_R \tilde{W}_R \tilde{W}_R + \text{h.c.}  \right).
\end{split}
\end{align}
\begin{itemize}
 \item Yukawa RGEs:
\end{itemize}
\begin{align*}
(4\pi)^2 \frac{d}{dt} \lambda_i & = \left( 8|\vec{\lambda}|^2 + 4 |\vec{\lambda^c}|^2  \right) \lambda_i
+\frac{15}{8}\vec{\alpha^c} \cdot \vec{\lambda}\, {\alpha_i^c}  - \left( \frac{15}{2}g_4^2 + 3g_L^2 +3g_R^2 \right) \lambda_i \\
(4\pi)^2 \frac{d}{dt} \lambda^c_i & = \left( 8|\vec{\lambda^c}|^2
+4|\vec{\lambda}|^2  +\frac{15}{8}\bar{\sigma}_c^2
+\frac{3}{4} a^2 \right)\lambda^c_i + \frac{15}{8} \vec{\alpha} \cdot\vec{\lambda^c} \,\alpha_i
 - \left( \frac{15}{2}g_4^2 + 3g_L^2 +3g_R^2 \right) \lambda^c_i \\
(4\pi)^2 \frac{d}{dt} \alpha_i &= \left( \frac{19}{4}|\vec{\alpha}|^2 + |\vec{\alpha^c}|^2
\right) \alpha_i  + 2 \vec{\lambda^c}\cdot \vec{\alpha}\,\lambda^c_i - \left( \frac{31}{2}g_4^2 + 3g_L^2 \right) \alpha_i \\
(4\pi)^2 \frac{d}{dt} \alpha_i &= \left( \frac{19}{4}|\vec{\alpha}|^2 + |\vec{\alpha^c}|^2
\right) \alpha_i  + 2 \vec{\lambda^c}\cdot \vec{\alpha}\,\lambda^c_i - \left( \frac{31}{2}g_4^2 + 3g_L^2 \right) \alpha_i \\
(4\pi)^2 \frac{d}{dt} \alpha^c_i &= \left(\frac{19}{4}|\vec{\alpha^c}|^2 
+ |\vec{\alpha}|^2  + \frac{15}{8} \sigma_c^2
+\frac{3}{4}a^2 \right) \alpha^c_i + 2 \vec{\lambda}\cdot \vec{\alpha^c}\,\lambda_i- \left( \frac{31}{2}g_4^2 + 3g_R^2 \right) \alpha^c_i \\
(4\pi)^2 \frac{d}{dt} a &= \left( \frac{7}{2} a^2 +
\frac{15}{8}(\sigma_c^2 + \bar{\sigma}_c^2)+ \frac{15}{8}|\vec{\alpha^c}|^2
+ 2|\vec{\lambda^c}|^2
 \right) a - \left( \frac{15}{2}g_4^2 + 7 g_R^2 \right) a \\
(4\pi)^2 \frac{d}{dt} \sigma_c &= \left( \frac{19}{4}\sigma_c^2 +
\bar{\sigma}_c^2 +
\frac{15}{8}|\vec{\alpha^c}|^2 +\frac{3}{4}a^2
\right) \sigma_c - \left( \frac{31}{2}g_4^2 + 3g_R^2 \right) \sigma_c \\
(4\pi)^2 \frac{d}{dt} \bar{\sigma}_c &= \left( \frac{19}{4}\bar{\sigma}_c^2 +
\sigma_c^2 +
2|\vec{\lambda^c}|^2+ \frac{3}{4}a^2
\right) \bar{\sigma}_c - \left( \frac{31}{2}g_4^2 + 3g_R^2 \right) \bar{\sigma}_c \\
\end{align*}

\begin{itemize}
 \item A-terms:
\end{itemize}

\begin{align}
\begin{split}
(4\pi)^2 \frac{d}{dt} A^\lambda_i &= 
\left(6|\vec{\lambda}|^2+4|\vec{\lambda^c}|^2 \right) A^\lambda_i  + \left(18 \vec{\lambda} \cdot \vec{A^\lambda} \right)+ 8 \vec{\lambda^c}\cdot \vec{A}^{\lambda^c}\lambda_i \nonumber \\
& + \left( \frac{15}{8}\vec{\alpha^c}\cdot \vec{A^\lambda} + \frac{15}{4}\vec{\lambda} \cdot\vec{A}^{\alpha^c} \right) \alpha^c_i
+ \frac{15}{2}g_4^2 \left(2 M_4 \lambda_i - A^\lambda_i \right) \nonumber \\
& + 3 g_L^2 \left(2 M_L \lambda_i - A^\lambda_i \right)
+ 3 g_R^2 \left(2 M_R \lambda_i - A^\lambda_i \right) \\
(4\pi)^2 \frac{d}{dt} A^{\lambda^c}_i &= 
\left(6|\vec{\lambda^c}|^2+4|\vec{\lambda}|^2+ \frac{15}{8}\bar{\sigma}_c^2+\frac{3}{4}a^2\right) A^{\lambda^c}_i 
 \nonumber\\
& + \left(18 \vec{\lambda^c}\cdot \vec{A}^{\lambda^c}+8 \vec{\lambda} \cdot \vec{A^\lambda} +\frac{15}{4}\bar{\sigma}_c A^{\bar{\sigma}_c }+\frac{3}{2}a A_a\right) \lambda^c_i \nonumber\\
& 
+ \left( \frac{15}{8}\vec{\alpha}\cdot \vec{A}^{\lambda^c} + \frac{15}{4}\vec{\lambda^c} \cdot\vec{A^{\alpha}} \right) \alpha_i  
 \nonumber\\
& + \frac{15}{2}g_4^2 \left(2 M_4 \lambda^c_i - A^{\lambda^c}_i \right)
+ 3 g_L^2 \left(2 M_L \lambda^c_i - A^{\lambda^c}_i \right) 
+ 3 g_R^2 \left(2 M_R \lambda^c_i - A^{\lambda^c}_i \right) \\
(4\pi)^2 \frac{d}{dt} A^\alpha_i &= 
\left(\frac{23}{8}|\vec{\alpha}|^2+|\vec{\alpha^c}|^2 \right) A^\alpha_i 
+ \left(\frac{91}{8} \vec{\alpha} \cdot \vec{A^\alpha} +2 \vec{\alpha^c}\cdot \vec{A}^{\alpha^c}\right) \alpha_i 
+ \left( 2 \vec{\lambda^c}\cdot \vec{A^\alpha} + 4 \vec{\alpha} \cdot\vec{A}^{\lambda^c} \right) \lambda^c_i
\nonumber\\
& + \frac{31}{2}g_4^2 \left(2 M_4 \alpha_i - A^\alpha_i \right)
+ 3 g_L^2 \left(2 M_L \alpha_i - A^\alpha_i \right) \\
(4\pi)^2 \frac{d}{dt} A^{\alpha^c}_i &= \left(\frac{23}{8}|\vec{\alpha^c}|^2+ |\vec{\alpha}|^2+ \frac{15}{8}{\sigma}_c^2+\frac{3}{4}a^2\right) A^{\alpha^c}_i 
 \nonumber\\
& + \left(\frac{91}{8} \vec{\alpha^c}\cdot \vec{A}^{\alpha^c}+ 2 \vec{\alpha} \cdot \vec{A^\alpha} +\frac{15}{4}\sigma_c A^{\sigma_c}+\frac{3}{2}a A_a\right) \alpha^c_i \nonumber\\
& + \left( 2\vec{\lambda}\cdot \vec{A}^{\alpha^c} + 4\vec{\alpha^c} \cdot\vec{A^{\lambda}} \right) \lambda_i  
\nonumber \\
& + \frac{31}{2}g_4^2 \left(2 M_4 \alpha^c_i - A^{\alpha^c}_i \right)
+ 3 g_R^2 \left(2 M_R \alpha^c_i - A^{\alpha^c}_i \right) \\
(4\pi)^2 \frac{d}{dt} A^a &= \left(\frac{21}{2} a^2 +  \frac{15}{8}(\sigma_c^2 + \bar{\sigma}_c^2)+ \frac{15}{8}|\vec{\alpha^c}|^2
+ 2|\vec{\lambda^c}|^2
 \right) A^a \nonumber \\
& + \left( \frac{15}{4}(\sigma_c A^{\sigma_c} + \bar{\sigma}_c A^{\bar{\sigma}_c})+ \frac{15}{4} \vec{\alpha^c} \cdot \vec{A}^{\alpha^c}
+ 4 \vec{\lambda^c} \cdot \vec{A}^{\lambda^c}
  \right) a \nonumber \\
& 
+ \frac{15}{2}g_4^2 \left(2 M_4 a - A^a \right)
+ 7 g_R^2 \left(2 M_R a - A^a \right) \\
(4\pi)^2 \frac{d}{dt} A^{\sigma_c} &= \left( \frac{57}{4}\sigma_c^2 + \bar{\sigma}_c^2+ \frac{15}{8}|\vec{\alpha^c}|^2
+ \frac{3}{4} a^2 \right) A^{\sigma_c} + \left( 2 \bar{\sigma}_c A^{\bar{\sigma}_c} + \frac{15}{4} \vec{\alpha^c} \cdot \vec{A}^{\alpha^c} 
+ \frac{3}{2} a A^a \right) \sigma_c 
\nonumber \\
& + \frac{31}{2}g_4^2 \left(2 M_4 \sigma_c - A^{\sigma_c} \right)
+ 3 g_R^2 \left(2 M_R \sigma_c - A^{\sigma_c} \right) \\
(4\pi)^2 \frac{d}{dt} A^{\bar{\sigma}_c} &= \left( \frac{57}{4}\bar{\sigma}_c^2 + \sigma_c^2+ 2|\vec{\lambda^c}|^2
+ \frac{3}{4} a^2 \right) A^{\bar{\sigma}_c} + \left( 2 \sigma_c A^{\sigma_c} + 4 \vec{\lambda^c} \cdot \vec{A}^{\lambda^c} 
+ \frac{3}{2} a A^a \right) \bar{\sigma}_c 
\nonumber \\
& + \frac{31}{2}g_4^2 \left(2 M_4 \bar{\sigma}_c - A^{\bar{\sigma}_c} \right)
+ 3 g_R^2 \left(2 M_R \bar{\sigma}_c - A^{\bar{\sigma}_c} \right) \\
\end{split}
\end{align}

\begin{itemize}
 \item Soft masses:
\end{itemize}

\begin{align}
\begin{split}
(4\pi)^2 \frac{d}{dt} (m^2_f)_{ij} &= \left(2 \lambda^c_i \lambda^c_k + \frac{15}{8} \alpha_i \alpha_k\right)
(m^2_f)_{kj} + (m^2_f)_{ik} \left(2 \lambda^c_k \lambda^c_j + \frac{15}{8} \alpha_k \alpha_j\right)
\nonumber\\& +4\left(m^2_{F^c} + m^2_h\right) \lambda^c_i \lambda^c_j  +\frac{15}{4}\left(m^2_{\bar{F}}+m^2_\phi \right)\alpha_i\alpha_j
\nonumber\\
&+4 A^{\lambda^c}_i A^{\lambda^c}_j + \frac{15}{4} A^\alpha_i A^\alpha_j
-15 g_4^2 M_4^2 - 6 g_L^2 M_L^2 \\
(4\pi)^2 \frac{d}{dt} (m^2_{f^c})_{ij} &= \left(2 \lambda_i \lambda_k + \frac{15}{8} \alpha^c_i \alpha^c_k\right)
(m^2_{f^c})_{kj} + (m^2_{f^c})_{ik} \left(2 \lambda_k \lambda_j + \frac{15}{8} \alpha^c_k \alpha^c_j\right)
\nonumber\\&+4\left(m^2_F + m^2_h\right) \lambda_i \lambda_j  +\frac{15}{4}\left(m^2_{\bar{F^c}}+m^2_\phi\right)\alpha^c_i\alpha^c_j
\nonumber\\&+4 A^\lambda_i A^\lambda_j + \frac{15}{8} A^{\alpha^c}_i A^{\alpha^c}_j
-15 g_4^2 M_4^2 - 6 g_R^2 M_R^2 \\
(4\pi)^2 \frac{d}{dt} m^2_h &= 8 \left( (|\vec{\lambda}|^2+ |\vec{\lambda^c}|^2 )m^2_h
+ (m^2_{f^c})_{ij} \lambda_i \lambda_j +  (m^2_f)_{ij} \lambda^c_i \lambda^c_j
+ m^2_F |\vec{\lambda}|^2   + m^2_{F^c} |\vec{\lambda^c}|^2 \right. \nonumber \\
&\left. + |\vec{A^\lambda}|^2
 +  |\vec{A}^{\lambda^c}|^2 \right) - 6 g_L^2 M_L^2- 6 g_R^2 M_R^2 \\
(4\pi)^2 \frac{d}{dt} m^2_\phi &=  \frac{15}{8}\left(\sum_i(\alpha^2_i + {\alpha^c_i}^2)m^2_\phi
+ (m^2_f)_{ij} \alpha_i \alpha_j +  (m^2_{f^c})_{ij} \alpha^c_i \alpha^c_j
+ m^2_{\bar{F}}  |\vec{\alpha}|^2   + m^2_{\bar{F}^c}
|\vec{\alpha^c}|^2  \right. \nonumber \\
&\left. + |\vec{A^\alpha}|^2 + |\vec{A}^{\alpha^c}|^2 \right)
- 32 g_4^2 M_4^2 \\
(4\pi)^2 \frac{d}{dt} m^2_F &= 4\left( |\vec{\lambda}|^2  m^2_F + (m^2_{f^c})_{ij} \lambda_i \lambda_j +  m^2_h |\vec{\lambda}|^2 +  |\vec{A^\lambda}|^2 \right)  -15 g_4^2 M_4^2 - 6 g_L^2 M_L^2 \\
(4\pi)^2 \frac{d}{dt} m^2_{\bar{F}} &=
\frac{15}{4}\left(|\vec{\alpha}|^2  m^2_{\bar{F}}
+  (m^2_f)_{ij} \alpha_i \alpha_j +  m^2_\phi |\vec{\alpha}|^2
+ |\vec{A^\alpha}|^2 \right)   -15 g_4^2 M_4^2 - 6 g_L^2 M_L^2 \\
(4\pi)^2 \frac{d}{dt} m^2_{F^c} &= \left(4 |\vec{\lambda^c}|^2 + \frac{3}{2}a^2 +
\frac{15}{4} \bar{\sigma_c}^2  \right) m^2_{F^c}
+ 4 (m^2_f)_{ij} \lambda^c_i \lambda^c_j + 4 m^2_h |\vec{\lambda^c}|^2
+ \frac{3}{2} (m^2_{X^c} + m^2_{\bar{F^c}}) a^2 \nonumber\\
&
+ \frac{15}{4}(m^2_\Phi + m^2_{\bar{F_c}^\prime}) \bar{\sigma_c}^2 
 + 4|\vec{A}^{\lambda^c}|^2 + \frac{3}{2} {A^a}^2+
\frac{15}{4} {A^{\bar{\sigma}_c}}^2 
  -15 g_4^2 M_4^2 - 6 g_R^2 M_R^2 \\
(4\pi)^2 \frac{d}{dt} m^2_{\bar{F}_c} &= \left(\frac{15}{4} |\vec{\alpha^c}|^2 + \frac{3}{2}a^2 +
\frac{15}{4}\sigma_c^2  \right) m^2_{\bar{F}^c}
+ \frac{15}{4} (m^2_{f^c})_{ij} \alpha^c_i \alpha^c_j + \frac{15}{4} m^2_\phi \vec|{\alpha^c}|^2\nonumber\\
&
+\frac{3}{2} (m^2_{X^c} + m^2_{F^c}) a^2 
+ \frac{15}{4} (m^2_\Phi + m^2_{F_c^\prime}) \bar{\sigma_c}^2  +  \frac{15}{4} |\vec{A}^{\alpha^c}|^2 + \frac{3}{2} {A^a}^2
+\frac{15}{4} {{A}^{\sigma^c}}^2  \nonumber\\
&-15 g_4^2 M_4^2 - 6 g_R^2 M_R^2 \\
(4\pi)^2 \frac{d}{dt} m^2_{F_c^\prime} &=
 \frac{15}{4} \left(\sigma_c^2 m^2_{F_c^\prime}
+ (m^2_\Phi + m^2_{\bar{F}_c}) \sigma_c^2
+ {A^{\sigma_c}}^2\right) -15 g_4^2 M_4^2 - 6 g_R^2 M_R^2 \\
(4\pi)^2 \frac{d}{dt} m^2_{\bar{F}_c^\prime} &=
\frac{15}{4} \left( \bar{\sigma_c}^2 m^2_{\bar{F}_c^\prime}
+ (m^2_\Phi + m^2_{F^c}) \bar{\sigma_c}^2
+ {A^{\bar{\sigma}_c}}^2 \right)-15 g_4^2 M_4^2 - 6 g_R^2 M_R^2 \\
(4\pi)^2 \frac{d}{dt} m^2_{X^c} &= \frac{3}{2}\left( a^2 m^2_{X^c}
+ (m^2_{F^c} + m^2_{\bar{F^c}}) a^2
+  {A^a}^2 \right)- 16 g_R^2 M_R^2 \\
(4\pi)^2 \frac{d}{dt} m^2_\Phi &= \frac{15}{8} \left( (\sigma_c^2
 +\bar{\sigma_c}^2) m^2_\Phi 
 +(m^2_{\bar{F^c}} + m^2_{F_c^\prime})\sigma_c^2 + (m^2_{F^c} + m^2_{\bar{F_c}^\prime})
\bar{\sigma_c}^2 + {A^{\sigma_c}}^2 +{A^{\bar{\sigma}_c}}^2 \right)\nonumber\\
&- 32 g_4^2 M_4^2
\end{split}
\end{align}

\subsection*{RGEs between $M_L$ and $M_R$}

Neglecting interactions with singlet neutrinos lighter than $M_R$, the superpotential between $M_L$ and $M_R$ is given by: 
\begin{align}
\begin{split}
W &= (\alpha_q^A)_i\, q_i \bar{Q} A_\phi + (\alpha_l^A)_i\, l_i \bar{L} A_\phi 
+ (\alpha_q^T)_i\, q_i \bar{L} \bar{T}_\phi + (\alpha_l^T)_i\, l_i \bar{Q} T_\phi
+ (\alpha_q^G)_i\,  G_\phi q_i \bar{Q} \\
& + \lambda^u_i\, u_i^c Q h_u + \lambda^d_i\, d_i^c Q h_d +
 \lambda^e_i\, e_i^c L h_d.
\label{eq:MLlag} 
\end{split}
\end{align}
The boundary conditions for the above Yukawas at $M_R$ read:
\begin{gather}
 \lambda^u = \lambda^d 
= \lambda^e = \lambda 
\nonumber \\
\sqrt{24}\, \alpha_q^A = -\frac{\sqrt{24}}{3}\alpha_l^A = \sqrt{2} \,\alpha_q^T =\sqrt{2}\,\alpha_l^T = \alpha_q^G = \alpha.  \nonumber
\end{gather}
The first and the second line of \Eq{MLlag} are decoupled: in particular the RGEs for the second line Yukawas and soft masses are MSSM-like.

\end{document}